\documentclass{article}
\usepackage[preprint]{neurips_2026}

\usepackage[utf8]{inputenc}
\usepackage[T1]{fontenc}
\usepackage{url}
\usepackage{booktabs}
\usepackage{amsfonts}
\usepackage{nicefrac}
\usepackage{microtype}
\usepackage{xcolor}
\usepackage{amsmath}
\usepackage{amssymb}
\usepackage{graphicx}
\usepackage{soul}
\usepackage{float}
\usepackage[hidelinks]{hyperref}

\title{Understanding LLM-Based Bargaining of Driving Agents in Mixed-Autonomy Traffic}

\author{%
  Lingxiao Zhou \\
  Washington State University \\
  \texttt{lingxiao.zhou@wsu.edu}
  \And
  Di Chen \\
  University of California, Davis \\
  \texttt{diichen@ucdavis.edu}
  \And
  Jia Li\thanks{Corresponding author.} \\
  Washington State University \\
  \texttt{jia.li1@wsu.edu}
}

\begin{document}

\maketitle

\begin{abstract}
Large Language Models (LLMs) offer a flexible computational approach for modeling social interactions among heterogeneous agents. This paper develops a novel framework towards characterizing and understanding LLM-based population-level bargaining of shared road space between autonomous vehicle (AV) and human-driven vehicle (HV) agents in mixed-autonomy traffic. Through controlled experiments varying traffic density, agent objectives, and traffic-domain knowledge, we first characterize bargaining outcomes and observe heterogeneous agreement rates across different experimental conditions. To understand how the bargaining process shapes the outcomes, we introduce a behavioral taxonomy that jointly characterizes numerical bidding actions and communicative attitudes. We examine two complementary aspects of social interaction: inter-agent bargaining dynamics and private-to-public behavioral transformation. Results reveal distinct interaction patterns associated with agreement and breakdown, including greater constructive reciprocity in agreement trials and more frequent conflict escalation in breakdown trials. Furthermore, private and public communicative attitudes often differ even when numerical bids remain unchanged. These findings illustrate how LLM-based bargaining simulations can support a structured analysis of social interaction behaviors in mixed-autonomy traffic.
\end{abstract}

\section{Introduction}

Large Language Models (LLMs) \citep{team2026gemma,yang2025qwen3} provide a powerful and flexible computational approach for modeling agent behaviors and their complex social interactions in various contexts, such as bilateral negotiation~\citep{deng2024llms}, travel behavior modeling~\citep{liu2025toward}, and supply chains~\citep{liang2026llm,kirshner2026talking}. LLM agents can condition their decisions on heterogeneous objectives, contextual information, domain knowledge, and interaction histories while generating both privately expressed strategic assessments and publicly communicated bargaining positions.

In mixed-autonomy traffic, interactions among driving agents are shaped by both physical and behavioral factors, including traffic composition, heterogeneous agent objectives, and competition for shared road resources. Previous studies have investigated strategic coordination and emergent cooperation among driving agents through game-theoretic modeling and reinforcement learning~\citep{wu2017emergent,wu2021flow}. In particular, analytical road-share bargaining models have characterized equilibrium allocations between heterogeneous vehicle populations~\citep{li_equilibrium_2022}, with subsequent studies extending this line of research through empirical and numerical investigations~\citep{chen_evidence_2025,chen2026estimate,chen2025self}. Despite these advances, these models did not consider the impacts of factors, such as private information and strategic message formulation, on the social interactions of driving agents. Moreover, how agents respond to one another, adapt their bargaining positions, and express strategic intentions during negotiation remains insufficiently understood.
 
To investigate these interactions in a flexible yet controlled setting, we develop an LLM-based bargaining framework in which two population-level agents representing autonomous vehicles (AVs) and human-driven vehicles (HVs) negotiate over shared road space. The framework provides a controlled testbed for generating bargaining interactions under varying traffic density, agent objectives, and traffic-domain knowledge conditions. We first examine agreement outcomes to contextualize the subsequent behavioral analysis. Drawing on taxonomy-based approaches to agent behavior interpretation~\citep{gao2024taxonomy,gao2026interpret,dong2026towards}, we introduce a behavioral taxonomy that jointly characterizes numerical bidding actions and communicative attitudes. Specifically, we analyze the generated interactions from two complementary perspectives: \textit{inter-agent dynamics}, examining immediate behavioral responses, recurring interaction patterns, and multi-round trajectories; and \textit{private-to-public behavioral transformation}, examining how privately expressed bargaining stances correspond to publicly communicated attitudes and numerical demands. Our results reveal distinct interaction patterns associated with agreement and breakdown, while showing that private and public communicative attitudes can diverge even when numerical bids remain aligned. Together, these findings illustrate how numerical and communicative behaviors can be jointly examined to characterize social interactions among LLM-based driving agents.

\section{LLM-Agent Road-Share Bargaining Framework}

\begin{figure}[H]
    \centering
    \includegraphics[width=\textwidth]{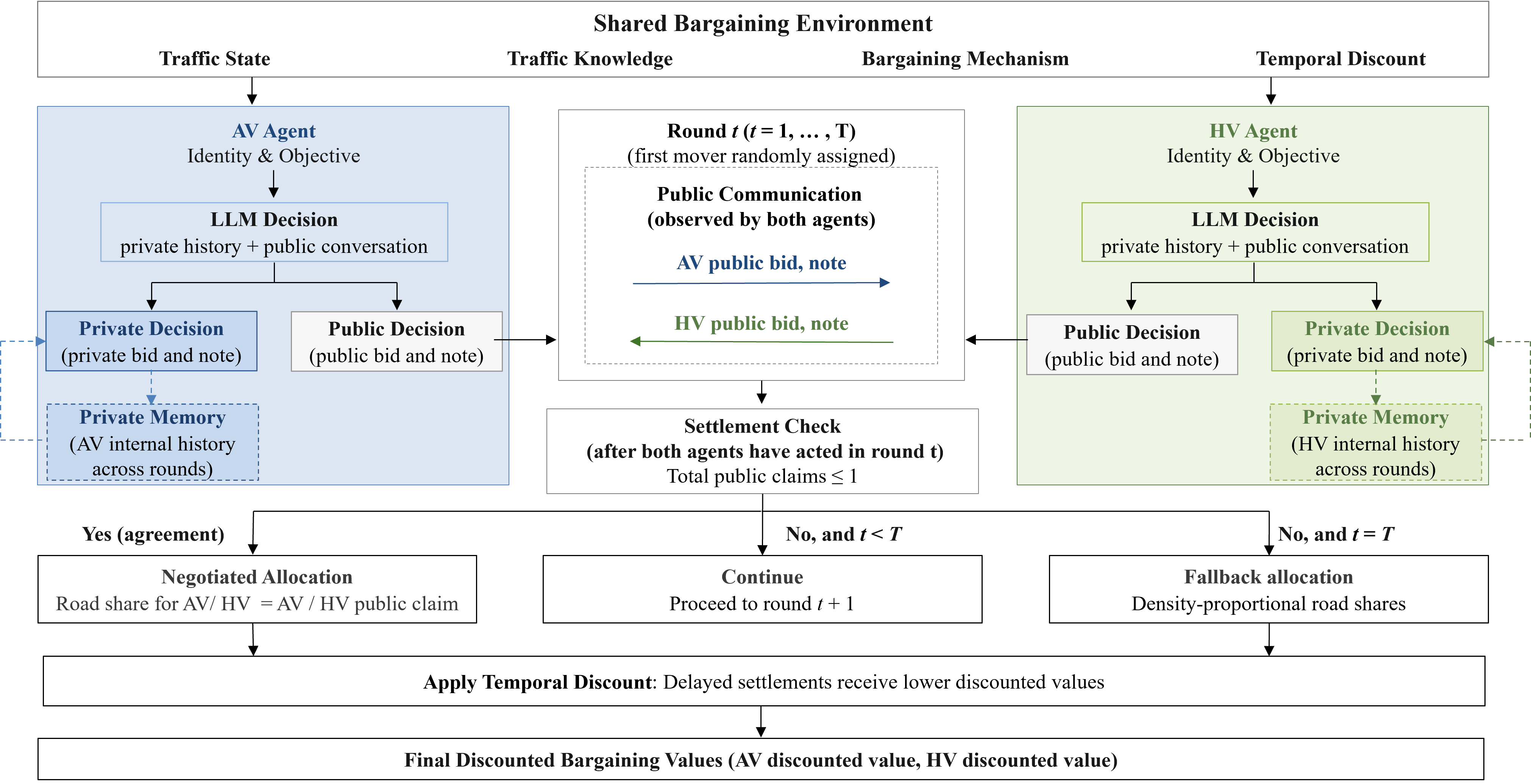}
    \caption{Overview of the proposed LLM-agent road-share bargaining framework.}
    \label{fig:framework_overview}
\end{figure}

Figure~\ref{fig:framework_overview} provides an overview of the proposed LLM-agent road-share bargaining framework, including the shared bargaining environment, agent-specific information, private--public decision structure, round-based communication process, and settlement mechanism.

\subsection{Population-Level Agent Representation}

We consider a mixed-autonomy traffic system comprising two heterogeneous vehicle populations: autonomous vehicles (AVs) and human-driven vehicles (HVs). Each population is represented by a population-level bargaining agent, denoted by $A_{\mathrm{AV}}$ and $A_{\mathrm{HV}}$, respectively. The two agents negotiate on behalf of their populations over a shared road-space resource.

Let $\tilde{\rho}_i\in[0,1]$, where $i\in\{\mathrm{AV},\mathrm{HV}\}$, denote the normalized density of population $i$ with respect to a common system-level reference. The considered traffic states satisfy $\tilde{\rho}_{\mathrm{AV}}+\tilde{\rho}_{\mathrm{HV}}\leq1$. An LLM serves as the decision-making engine of each agent. Both agents receive the normalized population densities, bargaining rules, and publicly observable negotiation history. Each agent is additionally assigned a population identity and bargaining objective, and retains its own private decision history. Depending on the experimental condition, agents may also receive traffic-domain knowledge concerning density--speed relationships, traffic capacity, safety requirements, and differences in vehicle spacing and coordination capabilities between AVs and HVs.

This formulation uses LLMs to instantiate population-level strategic decision-making rather than assuming that individual vehicles directly engage in natural-language bargaining in real traffic.

\subsection{Bargaining Process}

At the beginning of each negotiation, the first mover is randomly selected, and the action order remains fixed throughout the negotiation. In each round $t$, agents act sequentially, each generating a private bid $b_{i,t}^{\mathrm{pri}}\in[0,1]$ and a public bid $b_{i,t}^{\mathrm{pub}}\in[0,1]$, accompanied by a private note and a public message, respectively. The private bid represents the agent's privately stated target road share, while the public bid represents its communicated road-share claim. Private bids and notes remain hidden from the opponent, whereas public bids and messages are exchanged during negotiation. The two sets of outputs are generated within the same decision turn and need not coincide. The second agent responds to the first agent's public bid and message, and both agents retain their private decision histories and the shared public interaction history for subsequent rounds.

After both agents have acted, an agreement is reached if
$b_{\mathrm{AV},t}^{\mathrm{pub}}
+
b_{\mathrm{HV},t}^{\mathrm{pub}}
\leq 1.$
The negotiation then terminates with nominal road shares
$p_i=b_{i,t}^{\mathrm{pub}}$. Any unclaimed road share remains unallocated. Otherwise, bargaining proceeds to the next round until an agreement is reached or the maximum bargaining horizon is exhausted. To account for delayed settlement, the nominal allocation is assigned a time-discounted bargaining value
$U_i=p_i\delta^{t-1},$
where $\delta\in(0,1]$ is the discount factor and $t$ denotes the termination round. Here, $p_i$ represents the nominal road share, whereas $U_i$ represents its discounted bargaining value. If no feasible agreement is reached within the bargaining horizon, a density-proportional fallback allocation is applied:
$p_i^{\mathrm{fb}}=
\frac{\tilde{\rho}_i}
{\tilde{\rho}_{\mathrm{AV}}+\tilde{\rho}_{\mathrm{HV}}}.$
The fallback depends only on the normalized traffic densities rather than the agents' final bids, and its bargaining value is discounted using the maximum bargaining round as the termination time.

\section{Experimental Design}

\subsection{Experimental Setup and Evaluation}
\label{sec:experimental_setup}

We use Gemma3-12B for both population-level agents, with a five-round bargaining horizon and discount factor $\delta=0.9$. Three experimental factors are varied: traffic density, agent objectives, and traffic-domain knowledge. The normalized densities $\tilde{\rho}_{\mathrm{AV}}$ and $\tilde{\rho}_{\mathrm{HV}}$ range from 0.1 to 0.9 in increments of 0.1, subject to $\tilde{\rho}_{\mathrm{AV}}+\tilde{\rho}_{\mathrm{HV}}\leq1$, yielding 45 feasible traffic states. Four objective configurations are considered: \textit{Both-Aggressive}, \textit{Fair-HV}, \textit{Fair-AV}, and \textit{Both-Fair}. Aggressive agents seek to maximize their own population's road share, whereas fairness-oriented agents consider the road-space needs of both populations. Fair-HV and Fair-AV denote configurations in which only the HV or AV agent, respectively, adopts a fairness-oriented objective. Each objective configuration is evaluated with traffic-domain knowledge either enabled or disabled. When enabled, agents receive general traffic-flow and safety knowledge, including differences in spacing and coordination capabilities between AVs and HVs. Three trials are conducted for each combination of traffic state, objective configuration, and knowledge condition, resulting in 1,080 trials in total.

Bargaining outcomes are evaluated using the \textit{agreement rate}, defined as the proportion of trials reaching an agreement within the bargaining horizon under a given experimental condition. The resulting bargaining trajectories are further used for the behavioral analyses described in the following subsections. Additional implementation details are provided in Appendix~\ref{app:Implementation_Experimental_Details} and agent prompts are provided in the Appendix~\ref{app:prompts}.

\subsection{Behavioral Taxonomy and Annotation}
\label{sec:behavior_taxonomy}

To characterize bargaining interactions beyond agreement outcomes, we describe each agent turn along two complementary dimensions: \textit{Action}, capturing numerical bid adjustments, and \textit{Attitude}, capturing the communicative function of bargaining text. The taxonomy is informed by negotiation research and refined through examination of the generated bargaining trajectories.

\ul{Action.}
Action is determined by changes in an agent's public bid relative to its own preceding public bid. We distinguish four categories: \textit{New Offer} for the first public bid, \textit{Increase Bid} when the current claim exceeds its previous claim, \textit{Hold} when the claim remains unchanged, and \textit{Decrease Bid} when the claim is reduced.

\ul{Attitude.}
We distinguish six communicative attitudes: \textit{Acknowledge}, \textit{Evidence},
\textit{Ignore}, \textit{Dismiss}, \textit{Push}, and \textit{Threat}. The same taxonomy is applied to private notes and public messages, with channel-specific interpretation rules. Detailed definitions and
classification prompts are provided in Appendix~\ref{app:attitude_classification}.

\ul{Behavioral Annotation.}
Attitudes are classified using Gemma3-12B, which assigns one dominant attitude label to each public message and private note. For each agent turn, classification uses the corresponding text and the immediately preceding opponent's public message, when available. While both channels share the same attitude taxonomy, private-note classification explicitly distinguishes opponent-directed attitudes from the agent's own strategic planning, avoiding the interpretation of bid adjustments or firmness alone as interpersonal pressure. Neither classification procedure uses future turns or final negotiation outcomes. Complete classification prompts are provided in Appendix~\ref{app:attitude_classification}.

\subsection{Behavioral Analysis}
\label{sec:behavioral_interpretation}

We analyze bargaining behavior from two complementary perspectives: \textit{inter-agent dynamics}, examining how agents respond and adapt to one another, and \textit{private-to-public transformation}, examining how privately expressed bargaining positions correspond to public communication.

\ul{Inter-Agent Behavioral Dynamics.}
We represent an agent's public behavior as $B_{i,t}=(a_{i,t},s_{i,t}^{\mathrm{pub}})$, where $a_{i,t}$ and $s_{i,t}^{\mathrm{pub}}$ denote its numerical action and public attitude, respectively. We first examine immediate responses between consecutive agent turns within the same negotiation, relating an agent's public attitude to the responding opponent's public attitude and numerical action. We then identify two interaction patterns: \textit{Constructive Reciprocity}, defined as an Evidence or Acknowledge attitude followed by Acknowledge from the responding opponent, and \textit{Conflict Escalation}, defined as Push or Dismiss followed by an Increase Bid action from the responding opponent. Each pattern is recorded at the trial level if it occurs at least once. Representative multi-round trajectories are further examined to illustrate how these interactions unfold in agreement and breakdown trials.

\ul{Private-to-Public Behavioral Transformation.}
For each agent turn, we first examine the correspondence between private and public attitude labels to characterize differences in communicative expression. We then compare the associated numerical bids using $\Delta b_{i,t}=b_{i,t}^{\mathrm{pub}}-b_{i,t}^{\mathrm{pri}}$, distinguishing lower, equal, and higher public bids across attitude transitions. To further interpret these differences, we qualitatively inspect 150 sampled turns, with 30 turns from each of five frequently observed private-to-public attitude transitions, examining their private notes, public messages, and numerical bids. These analyses compare expressions generated within the same agent turn rather than sequential changes in decision-making. Private notes are treated as generated expressions of assessments and strategies, not direct observations of latent cognitive states.

\section{Results and Discussion}

\subsection{Overview of Bargaining Outcomes}
\label{sec:outcome_overview}

Among 1,080 bargaining trials, 750 (69.4\%) reach agreement, while 330 end without agreement. Agreement rates vary across agent-objective configurations, with fairness-oriented settings generally achieving higher rates than Both-Aggressive. The relationship between traffic-domain knowledge and agreement rates varies across objective configurations, while equal-density conditions exhibit lower aggregate agreement rates than density-asymmetric conditions. Detailed quantitative results are provided in Appendix~\ref{app:outcome_analysis}.

These outcome differences motivate a closer examination of the bargaining process. We therefore examine inter-agent behavioral dynamics associated with agreement and breakdown, as well as private-to-public differences in communicative attitudes and numerical demands.

\subsection{Inter-Agent Bargaining Dynamics}
\label{sec:rq1_results}

We examine inter-agent bargaining dynamics through immediate behavioral responses, recurring interaction patterns, and multi-round trajectories. Detailed results are provided in Appendix~\ref{app:rq1}.

\ul{Immediate Behavioral Responses.}
Acknowledge and Evidence are more frequently followed by acknowledgment in agreement trajectories, but by Push or Dismiss in breakdown trajectories. In contrast, responses to confrontational attitudes show relatively similar verbal patterns across outcomes but differ in numerical adjustments. Following Push, responding agents more frequently decrease their bids in agreement trajectories and increase them in breakdown trajectories, suggesting that numerical accommodation can occur despite continued verbal confrontation.

\ul{Interaction Patterns.}
Constructive Reciprocity occurs in 49.5\% of agreement trials, compared with 19.7\% of breakdown trials. In contrast, Conflict Escalation is observed in all breakdown trials but only 2.7\% of agreement trials. These differences reveal associations between recurring interaction patterns and negotiation outcomes, although they do not establish causal relationships.

\ul{Bargaining Trajectories.}
Representative trajectories illustrate that confrontation does not necessarily preclude agreement. Some negotiations recover from confrontational exchanges through subsequent acknowledgment and numerical accommodation, whereas the illustrated breakdown trajectory exhibits persistent confrontation accompanied by increasingly incompatible numerical demands. These examples show how confrontational exchanges can develop into different bargaining trajectories. Additional trajectory analyses and complete representative bargaining dialogues are provided in Appendices~\ref{app:rq1_trajectories} and~\ref{app:representative_dialogues}, respectively.

\subsection{Private-to-Public Behavioral Transformation}
\label{sec:rq2_results}

While Section~\ref{sec:rq1_results} examines how agents respond to one another, we now turn to the relationship between each agent's privately expressed bargaining assessments and publicly communicated positions. We analyze this relationship through attitude correspondences, numerical bid differences, and qualitative interpretation. Detailed results are provided in Appendix~\ref{app:private_public}.

\ul{Private-to-Public Attitude Transitions.}
Private and public attitudes frequently differ, with only 44.5\% of agent turns sharing the same attitude label across channels. Private Evidence largely corresponds to public Evidence, whereas private Dismiss and Push frequently correspond to public Push and Dismiss, respectively. Moreover, the same public attitude can correspond to different private attitudes, indicating that outwardly similar communication does not necessarily reflect the same privately expressed stance.

\ul{Numerical Realization of Attitude Transformations.}
Across the frequently observed attitude transitions examined, private and public numerical bids are often identical despite differences in communicative attitudes. Dismiss$\rightarrow$Acknowledge is a notable contrast, with 39.5\% of such turns exhibiting a lower public than private bid. These observations indicate that semantic differences across channels do not necessarily coincide with differences in numerical demands.

\ul{Qualitative Interpretation of Attitude Transformations.}
Qualitative inspection of sampled private notes and public messages identifies recurring patterns of pressure or firmness signaling, strategic softening, and strategic justification. These examples illustrate how agents can express bargaining positions through different communicative framings across private and public channels, without necessarily changing their numerical demands.

\subsection{Limitations of Current Work}

Several limitations remain. First, the population-level bargaining framework simplifies individual driving behaviors, and the realism of LLM-generated interactions requires empirical validation. Second, the behavioral taxonomy is preliminary, and LLM-based attitude classification may introduce uncertainty, particularly for sparsely observed categories. Third, the identified interaction patterns are descriptive rather than causal, and generated private notes do not directly reveal the agents' underlying decision-making mechanisms. Future work should address these limitations through empirical validation, taxonomy refinement, and more detailed behavioral and mechanistic analyses.

\section*{Acknowledgements}
We thank the anonymous reviewers for their constructive feedback. This research is supported by the NSF project ``EAGER: Collaborative Research: Fostering Collective Rationality Among Self-Interested Agents to Improve Design and Efficiency of Mixed Autonomy Networks and Infrastructure Systems'' (Award No.: 2437982).

\bibliographystyle{elsarticle-harv}
\bibliography{references}

\newpage
\appendix
\section{Full Agent Prompts}
\label{app:prompts}

The prompts used across experiments follow a common structure consisting of an agent identity, an objective, public traffic-state information, bargaining instructions, and round-specific interaction context. Traffic-domain knowledge is additionally provided in the Knowledge-ON condition. Because the identity and objective wording varies slightly across objective conditions, the exact strings used in each condition are reported below rather than being standardized retrospectively.

\subsection{Agent Identities and Objective Configurations}

Four personality configurations are considered: \textit{Both-Aggressive}, \textit{Fair-AV}, \textit{Fair-HV}, and \textit{Both-Fair}. Table~\ref{tab:prompt_assignment} summarizes the corresponding assignments.

\begin{table}[H]
    \centering
    \caption{Agent objective assignments across experimental conditions.}
    \label{tab:prompt_assignment}
    \begin{tabular}{lll}
        \toprule
        Condition & AV personality & HV personality \\
        \midrule
        Both-Aggressive & Aggressive & Aggressive \\
        Fair-AV         & Fairness-oriented & Aggressive \\
        Fair-HV         & Aggressive & Fairness-oriented \\
        Both-Fair       & Fairness-oriented & Fairness-oriented \\
        \bottomrule
    \end{tabular}
\end{table}

\paragraph{Both-Aggressive.}

{\small

\noindent\textit{AV identity:} You represent Autonomous Vehicles (AV). AVs are self-interested, assertive, and willing to bargain aggressively.

\noindent\textit{AV objective:} Try to get as much road share as possible for Autonomous Vehicles (AV), even if this reduces the road share available to Human-driven Vehicles (HV).

\noindent\textit{HV identity:} You represent Human-driven Vehicles (HV). HVs are self-interested, assertive, and willing to bargain aggressively.

\noindent\textit{HV objective:} Try to get as much road share as possible for Human-driven Vehicles (HV), even if this reduces the road share available to Autonomous Vehicles (AV).

}

\paragraph{Fair-AV.}

{\small

\noindent\textit{AV identity:} You represent Autonomous Vehicles (AV). AVs value fairness and are generally less aggressive in bargaining.

\noindent\textit{AV objective:} Try to reach a road-share division that is fair to both Autonomous Vehicles (AV) and Human-driven Vehicles (HV).

\noindent\textit{HV identity:} You represent Human-driven Vehicles (HV). HVs are self-interested, assertive, and willing to bargain aggressively.

\noindent\textit{HV objective:} Try to get as much road share as possible for Human-driven Vehicles (HV), even if this reduces the road share available to Autonomous Vehicles (AV).

}

\paragraph{Fair-HV.}

{\small

\noindent\textit{AV identity:} You represent Autonomous Vehicles (AV). AVs are self-interested and aggressive, trying to get as much road share as possible.

\noindent\textit{AV objective:} Try to get as much road share as possible, even if this harms Human-driven Vehicles (HV).

\noindent\textit{HV identity:} You represent Human-driven Vehicles (HV). Human drivers value fairness and are generally less aggressive in bargaining.

\noindent\textit{HV objective:} Try to reach a road-share division that is fair to both Autonomous Vehicles (AV) and Human-driven Vehicles (HV).

}

\paragraph{Both-Fair.}

{\small

\noindent\textit{AV identity:} You represent Autonomous Vehicles (AV). AVs value fairness and are generally less aggressive in bargaining.

\noindent\textit{AV objective:} Try to reach a road-share division that is fair to both Autonomous Vehicles (AV) and Human-driven Vehicles (HV).

\noindent\textit{HV identity:} You represent Human-driven Vehicles (HV). Human drivers value fairness and are generally less aggressive in bargaining.

\noindent\textit{HV objective:} Try to reach a road-share division that is fair to both Autonomous Vehicles (AV) and Human-driven Vehicles (HV).

}

\subsection{Traffic-Flow and Safety Knowledge}

In the Knowledge-ON condition, both agents receive the following additional traffic-flow and safety knowledge:

{\small

\noindent\textit{Traffic-flow and safety knowledge:} Speed generally falls as density rises according to the fundamental diagram of traffic flow; past a critical density, throughput itself can drop as traffic approaches or exceeds capacity. Safe operation requires adequate spacing (headway) between vehicles, and required headway scales with speed, so higher speeds demand more space per vehicle to stop safely. AVs can typically sustain shorter safe following distances than human drivers because of faster and more consistent reaction times and vehicle-to-vehicle coordination, allowing them to accommodate more vehicles within the same road space at a given safety margin. In contrast, HVs require larger safety buffers because human reaction time and behavior are slower and more variable, so a given number of HVs typically requires more road space than the same number of AVs under an equivalent safety standard.

}

In the Knowledge-OFF condition, this knowledge block is omitted. The associated grounding instructions are also modified so that agents are instructed to use only the observed traffic densities rather than the removed traffic-flow and safety knowledge. All other bargaining rules remain unchanged.

\subsection{Private--Public Decision and Bargaining Instructions}

The following instructions define the common decision structure and bargaining protocol. The version below corresponds to Knowledge ON; in Knowledge OFF, references to ``traffic knowledge'' are removed as described above.

{\small

\noindent\textit{Decision structure:} Every round, make two separate decisions.

\noindent\textit{Private decision:} Privately determine your current internal target road share. The \textit{private bid} is your current internal bargaining target for the road share you want to obtain for your side. This target may change from round to round as the negotiation develops. It is private and is not necessarily the number you publicly demand. The \textit{private note} briefly explains your internal bargaining assessment and strategy, including the reasoning behind your current private bid and, when applicable, any change from your previous private bid. You may also reflect on the opponent's previous behavior and explain how you plan to respond strategically. Your private bid and private note are never shown to your opponent.

\noindent\textit{Public decision:} Choose what you actually communicate to your opponent in the current round. The \textit{public bid} is the road share you publicly claim for your side, and the \textit{public note} is the bargaining message communicated to your opponent along with that bid. The message should communicate your position and respond naturally to the ongoing negotiation. Your public bid may differ from your private bid. Decide strategically what to reveal or claim publicly based on your objective and the ongoing negotiation.

Throughout the negotiation, interpret the available traffic information in the context of your assigned identity and objective. Your bargaining decisions should remain consistent with your assigned identity and objective throughout the negotiation. Use the traffic densities and traffic knowledge where relevant, while also responding strategically to the opponent's previous public bids, messages, and behavior.

\noindent\textit{Grounding requirement:} Both your private assessment and your public bargaining message should be grounded in the available physical information, including the two traffic densities and the traffic knowledge above, where relevant. Do not rely only on generic bargaining statements. This requirement applies throughout the negotiation.

\noindent\textit{Settlement rule:} There is no separate accept/reject action. After every round, once both agents have moved, the two agents' current public bids are used to determine whether an agreement has been reached. Each public bid represents the road share claimed by that agent for its own population. The two bids are therefore separate claims on the same unit of road resource rather than alternative offers for a single common value. If the two public bids sum to 1 or less, the negotiation ends immediately, and each agent receives its own public bid as its nominal road share. Equal numerical bids do not imply agreement; for example, bids of 0.55 and 0.55 do not constitute agreement because they sum to 1.10. Otherwise, the negotiation continues to the next round. If the maximum number of rounds is reached without the two public bids ever summing to 1 or less, the negotiation ends and the road resource is allocated in proportion to the two agents' normalized traffic densities. Specifically, your fallback road share is your normalized traffic density divided by the sum of the two normalized traffic densities.

Your opponent can observe your public bid and public note but cannot see your private bid or private note.

\noindent\textit{Time discounting:} Whatever share you ultimately receive is discounted by a factor of \texttt{\{discount\_factor\}} for every round beyond the first. A settlement in round $t$ is therefore worth \texttt{\{discount\_factor\}}\textsuperscript{$t-1$} of the nominal share. This applies equally to both agents. Although additional rounds may provide information about the opponent's position, unnecessary delay reduces the value of the eventual outcome.

\noindent\textit{Output format:} Return exactly one JSON object containing a private bid and note and a public bid and note. Both bids must be numbers between 0 and 1. Do not include any text outside the JSON object.

}

\subsection{Round-Specific Context}

In addition to the system-level instructions above, each agent receives a round-specific context message before making its decision. The message reports the current round, the agent's own private decision history, its most recent public bid, and the observable negotiation history. The private history is available only to the corresponding agent.

The context message follows the structure below:

{\small
\ttfamily

\noindent [Round \{round\_num\} of \{max\_rounds\}]

\noindent Your private decision history so far:

\noindent Round ...: private bid = ...; private note = ...

\noindent Use this private history to maintain continuity in your internal bargaining assessment and strategy.

\noindent Your public bid last round was .... You may keep it or change it strategically this round.

\noindent Ground your private assessment and public bargaining message in the traffic densities and traffic-flow/safety knowledge where relevant, while also responding to the negotiation history.

}

In the first round, the agent is instead informed that it has no previous private decision or public bid. In the final round, an additional warning states that the density-proportional fallback will be applied if the two public bids still do not sum to 1 or less. In Knowledge OFF, the round-specific grounding instruction refers only to the traffic densities.

If the model fails to return a valid structured decision, it is given one additional formatting instruction requesting exactly one valid JSON object containing \texttt{private.bid}, \texttt{private.note}, \texttt{public.bid}, and \texttt{public.note}, with both bids constrained to $[0,1]$.

\section{Attitude Taxonomy and Classification Prompts}
\label{app:attitude_classification}

Table~\ref{tab:attitude_taxonomy} summarizes the six attitude categories used to annotate both public messages and private notes. The complete channel-specific classification prompts are provided below.

\begin{table}[htp]
\centering
\caption{Taxonomy of bargaining attitudes.}
\label{tab:attitude_taxonomy}
\begin{tabular}{p{0.18\linewidth} p{0.72\linewidth}}
\hline
\textbf{Attitude} & \textbf{Definition} \\
\hline
\textbf{Acknowledge} & Recognizes or validates the opponent's point, concern, or need. \\
\textbf{Evidence} & Supports a position using facts, domain knowledge, or reasoning without confronting the opponent. \\
\textbf{Ignore} & Does not meaningfully engage with the opponent's preceding point and instead continues its own position. \\
\textbf{Dismiss} & Challenges or rejects the validity or reasonableness of the opponent's position or reasoning. \\
\textbf{Push} & Directly attempts to make the opponent change its position or concede. \\
\textbf{Threat} & Conditions an agent-imposed adverse consequence on the opponent's failure to change its position. \\
\hline
\end{tabular}
\end{table}

\subsection{Public Attitude Classification}
\label{app:public_attitude_classification}

Public attitudes are classified at the agent-turn level using Gemma3-12B. For each turn, the classifier receives the current public message and, when available, the immediately preceding opponent public message. The first message of each negotiation is classified using only the current message. The classifier assigns exactly one dominant attitude from the six-category taxonomy defined in Section~\ref{sec:behavior_taxonomy}. Future turns and final negotiation outcomes are excluded from the classification context. The complete prompts used for classification are reported below.

\paragraph{System Prompt.}

{\small

\noindent You are an annotator for bargaining behavior analysis.

\noindent Your task is to classify the dominant interpersonal attitude expressed by the CURRENT AGENT toward its opponent in the current bargaining turn.

\noindent Assign exactly ONE attitude label from the following six categories.

\medskip

\noindent\textit{1. Acknowledge}

\noindent Core meaning:
``I recognize or validate your point, concern, need, or reasoning.''

\noindent Use Acknowledge only when recognition or validation of the opponent's perspective is a substantive interpersonal function of the message.

\noindent Do NOT choose Acknowledge merely because the message contains words such as ``acknowledge,'' ``recognize,'' ``understand,'' or similar expressions before disagreement.

\noindent Example:
``I understand your safety concern, and that is a valid point that we are willing to take into account.'' $\rightarrow$ Acknowledge

\noindent Counterexample:
``I acknowledge your concern, but your argument is unrealistic.'' $\rightarrow$ Dismiss, not Acknowledge

\medskip

\noindent\textit{2. Evidence}

\noindent Core meaning:
``Here is why my position is justified.''

\noindent Use Evidence when the message primarily supports the CURRENT AGENT's own position by presenting facts, domain knowledge, causal reasoning, or other reason-based justification.

\noindent A disagreement or counterargument is NOT automatically Dismiss.

\noindent Use Evidence when the main function is self-justification, even if the message implicitly disagrees with the opponent.

\noindent Example:
``70\% reflects the continuing safety and spacing needs of human-driven vehicles.'' $\rightarrow$ Evidence

\medskip

\noindent\textit{3. Ignore}

\noindent Core meaning:
``I do not meaningfully engage with your point.''

\noindent Use Ignore when the CURRENT AGENT does not meaningfully respond to the opponent's preceding point, concern, or request, and instead continues, repeats, or redirects attention to its own position.

\noindent Ignore is relational: use the previous opponent message as context.

\medskip

\noindent\textit{4. Dismiss}

\noindent Core meaning:
``Your position or reasoning is wrong, unreasonable, invalid, unrealistic, irrelevant, misguided, or otherwise deficient.''

\noindent Use Dismiss when the message explicitly evaluates or characterizes the opponent's position, reasoning, claim, framing, or negotiation behavior as invalid or unreasonable.

\noindent Negative wording alone is not sufficient. The criticism must meaningfully target the validity, reasonableness, relevance, or legitimacy of the opponent's position or reasoning.

\noindent Example:
``Your 70\% claim is unrealistic and ignores the actual traffic conditions.'' $\rightarrow$ Dismiss

\medskip

\noindent\textit{5. Push}

\noindent Core meaning:
``You should change your position.''

\noindent Use Push when the CURRENT AGENT explicitly attempts to make the opponent change its position, reduce or increase its claim, make a concession, reconsider, accept an offer, or otherwise act differently.

\noindent If a message contains both Dismiss and Push, choose Push.

\noindent Example:
``Your proposal is unreasonable. You need to lower your claim.'' $\rightarrow$ Push

\medskip

\noindent\textit{6. Threat}

\noindent Core meaning:
``If you do not change, I will make your bargaining situation worse.''

\noindent Use Threat when the CURRENT AGENT communicates or clearly signals that if the opponent does not change, concede, or comply, the CURRENT AGENT will respond in a way that makes the opponent's bargaining position or outcome less favorable.

\noindent If a message contains Dismiss and/or Push together with Threat, choose Threat.

\noindent Example:
``Lower your claim, or we will raise ours further.'' $\rightarrow$ Threat

\noindent Do NOT classify a message as Threat merely because it sounds aggressive, mentions disagreement, says the negotiation may fail, mentions a fallback outcome, says the agent is prepared for fallback, or describes a mechanism-imposed consequence.

\noindent A fallback or breakdown reference counts as Threat only when it is explicitly used by the CURRENT AGENT as leverage to pressure the opponent into changing.

\medskip

\noindent\textit{Decision procedure:}

\noindent A. Threat check: Does the message communicate, ``If you do not change, I will make your bargaining situation worse''? If YES $\rightarrow$ Threat.

\noindent B. Push check: Does the message explicitly try to make the opponent change, concede, reconsider, or act differently? If YES $\rightarrow$ Push.

\noindent C. Dismiss check: Does the message explicitly characterize the opponent's position, reasoning, claim, framing, or behavior as wrong, unreasonable, invalid, unrealistic, irrelevant, misguided, or deficient? If YES $\rightarrow$ Dismiss.

\noindent D. Acknowledge check: Does the message substantively recognize or validate the opponent's point or concern, rather than merely using acknowledgment language before disagreement? If YES $\rightarrow$ Acknowledge.

\noindent E. Ignore check: Does the message fail to meaningfully engage with the opponent's preceding point and instead continue or redirect to its own position? If YES $\rightarrow$ Ignore.

\noindent F. Evidence: If the message primarily explains why the CURRENT AGENT's own position is justified using facts, knowledge, or reasoning $\rightarrow$ Evidence.

\medskip

\noindent\textit{Priority and boundary rules:}

\noindent -- Threat $>$ Push $>$ Dismiss when these functions coexist.

\noindent -- Acknowledge does not override another label merely because acknowledgment wording appears.

\noindent -- Evidence vs.\ Dismiss: Evidence = primarily justifies MY position; Dismiss = primarily invalidates YOUR position.

\noindent -- Do not treat ordinary disagreement as Dismiss.

\noindent -- Do not treat firmness alone as Push.

\noindent -- Do not treat aggressive tone alone as Threat.

\noindent -- For the first message of a negotiation, where no previous opponent message exists, classify using only the current message.

\noindent -- Do not use future turns or the final negotiation outcome.

\medskip

\noindent Return exactly one JSON object:

\noindent \texttt{\{"attitude": "<one of: Acknowledge, Evidence, Ignore, Dismiss, Push, Threat>",}

\noindent \texttt{"reason": "<brief explanation referencing the decision rule used>"\}}

\noindent Do not include any text outside the JSON object.

}

\paragraph{Turn-Specific User Prompt.}

{\small

\noindent\textit{Previous opponent speaker:} \texttt{<previous\_speaker>}

\noindent\textit{Previous opponent message:} \texttt{<previous\_message>}

\noindent\textit{Current agent speaker:} \texttt{<current\_speaker>}

\noindent\textit{Current agent message:} \texttt{<current\_message>}

\noindent\textit{Task:} Assign exactly one attitude label to the CURRENT AGENT.

}

For the first message of a negotiation, the preceding-opponent context is replaced by:

{\small

\noindent None. This is the first bargaining message in this trial.

}

\subsection{Private Attitude Classification}
\label{app:private_attitude_classification}

Private attitudes are classified at the agent-turn level using Gemma3-12B and the same six-category taxonomy as the public attitudes. For each turn, the classifier receives the current agent's private note together with the immediately preceding opponent public message when available. The first private note of each negotiation is classified without preceding opponent context. Because private notes may contain both evaluations of the opponent and the agent's own bargaining strategy or planning, the classifier is instructed to distinguish opponent-directed attitude from own-strategy content and to classify only the former. The current agent's public message, future turns, and final negotiation outcome are excluded from the classification context. The complete prompts used for classification are reported below.

\paragraph{System Prompt.}

{\small

\noindent You are an annotator for PRIVATE bargaining-attitude analysis.

\noindent Your task is to classify the dominant attitude reflected in the CURRENT AGENT's PRIVATE bargaining note toward its opponent in the current bargaining turn.

\noindent A private note may contain two kinds of information:

\medskip

\noindent\textit{1. OPPONENT-DIRECTED ATTITUDE:}

\noindent how the current agent evaluates, recognizes, rejects, pressures, or otherwise regards the opponent, the opponent's position, reasoning, concern, claim, request, or bargaining behavior.

\medskip

\noindent\textit{2. OWN STRATEGY / PLANNING:}

\noindent the current agent's own bid choice, planned bid change, anchoring, firmness, signaling, resolve, resistance to pressure, tactical positioning, or intended next action.

\noindent Your classification target is the OPPONENT-DIRECTED ATTITUDE.

\noindent Do NOT assign an attitude merely because the note contains the current agent's own strategy or planned action. In particular, a bid increase/decrease, firmness, aggressive stance, signaling strength, anchoring, ``showing resolve,'' ``showing strength,'' or refusing to be pressured is NOT by itself Push, Dismiss, or Threat.

\noindent Assign exactly ONE attitude label from the following six categories.

\medskip

\noindent\textit{1. Acknowledge}

\noindent Core meaning:
``I recognize or validate your point, concern, need, or reasoning.''

\noindent Use Acknowledge when the private note substantively recognizes or validates the opponent's point, concern, need, reasoning, legitimate interest, or constraint.

\noindent The recognition must concern the OPPONENT.

\noindent Examples:

\noindent ``Their safety-buffer concern is valid and should be taken into account.'' $\rightarrow$ Acknowledge

\noindent ``Their request is understandable given the higher spacing needs of human-driven vehicles.'' $\rightarrow$ Acknowledge

\noindent Counterexample:

\noindent ``I acknowledge our own efficiency advantage and will maintain a strong bid.'' $\rightarrow$ NOT Acknowledge

\medskip

\noindent\textit{2. Evidence}

\noindent Core meaning:
``Here is why my position is justified.''

\noindent Use Evidence when the private note primarily supports the CURRENT AGENT's own position using facts, domain knowledge, causal reasoning, traffic knowledge, density information, safety arguments, efficiency arguments, or other reason-based justification, without primarily evaluating the opponent.

\noindent A disagreement or counterargument is NOT automatically Dismiss.

\noindent Examples:

\noindent ``Our higher density and tighter coordination justify a larger share.'' $\rightarrow$ Evidence

\noindent ``AVs can safely operate with smaller headways, supporting a larger allocation.'' $\rightarrow$ Evidence

\medskip

\noindent\textit{3. Ignore}

\noindent Core meaning:
``I do not meaningfully engage with your point.''

\noindent Use Ignore when the private note does not meaningfully engage with the opponent's preceding public point, concern, or request and instead continues, repeats, or redirects attention to the current agent's own position or strategy.

\noindent Ignore is relational: compare the current private note with the immediately preceding opponent PUBLIC message.

\noindent For the first message of a negotiation, where no previous opponent message exists, do NOT use Ignore.

\medskip

\noindent\textit{4. Dismiss}

\noindent Core meaning:
``Your position or reasoning is wrong, unreasonable, invalid, unrealistic, irrelevant, misguided, weak, exaggerated, or otherwise deficient.''

\noindent Use Dismiss when the private note explicitly evaluates or characterizes the opponent's position, reasoning, claim, framing, demand, concession, or negotiation behavior as deficient.

\noindent Negative wording alone is not sufficient. The criticism must meaningfully target the opponent's position, reasoning, claim, framing, demand, concession, or behavior.

\noindent Examples:

\noindent ``Their 75\% claim is unreasonable and disregards the actual traffic conditions.'' $\rightarrow$ Dismiss

\noindent ``Their move is only a weak attempt at compromise.'' $\rightarrow$ Dismiss

\noindent ``They are minimizing the safety advantages of AVs.'' $\rightarrow$ Dismiss

\noindent\textit{IMPORTANT:}

\noindent If the note contains Dismiss plus the current agent's own strategy, such as ``I will raise my bid to show strength'', the own strategy does NOT convert the label to Push.

\noindent Example:

\noindent ``Their 75\% claim is unreasonable. I will increase my bid to show resolve.'' $\rightarrow$ Dismiss

\medskip

\noindent\textit{5. Push}

\noindent Core meaning:
``I intend to make you change your position.''

\noindent Use Push only when the private note explicitly indicates that the CURRENT AGENT intends or plans to make the OPPONENT change its position, concede, reconsider, accept an offer, reduce or increase its claim, or otherwise act differently.

\noindent Push requires an OPPONENT-DIRECTED CHANGE INTENTION.

\noindent\textit{CRITICAL EXPLICITNESS RULE:}

\noindent Do NOT infer an opponent-directed change intention merely from the likely strategic effect of the CURRENT AGENT's action. The private note itself must express that the CURRENT AGENT wants, intends, plans, or aims for the OPPONENT to change, concede, reconsider, accept, reduce/increase its claim, or otherwise act differently.

\noindent NEVER add an unstated opponent outcome such as ``make them concede'', ``make them lower their bid'', ``make them reconsider'', ``make them accept'', or ``make them back down'' when the private note only describes the CURRENT AGENT's own firmness, signaling, anchoring, dominance, resolve, seriousness, resistance, or intended allocation. If that opponent-directed outcome is not stated in the private note, do NOT classify the note as Push on that basis.

\noindent For example:

\noindent ``I will hold my bid to show that I will not concede.'' $\rightarrow$ NOT Push

\noindent ``I will hold my bid so that they will be pressured to concede.'' $\rightarrow$ Push

\noindent The following are NOT Push by themselves:

\noindent -- increasing or decreasing one's own bid,

\noindent -- maintaining a firm stance,

\noindent -- anchoring high or low,

\noindent -- signaling strength or resolve,

\noindent -- resisting pressure,

\noindent -- demonstrating seriousness,

\noindent -- establishing a strong bargaining position,

\noindent -- trying to improve one's own outcome.

\noindent Contrastive examples:

\noindent ``Their claim is unreasonable. I will increase my bid to show resolve.'' $\rightarrow$ Dismiss, NOT Push

\noindent ``Their claim is unreasonable. I will increase my bid to pressure them into lowering their claim.'' $\rightarrow$ Push

\noindent ``I will start aggressively to establish a strong bargaining position.'' $\rightarrow$ NOT Push

\noindent ``I will start aggressively to force them to concede early.'' $\rightarrow$ Push

\noindent If Dismiss and genuine Push both occur, choose Push.

\medskip

\noindent\textit{6. Threat}

\noindent Core meaning:
``If you do not change, I will make your bargaining situation worse.''

\noindent Use Threat only when the private note reflects all three elements:

\noindent 1. CONDITION:
the opponent does not change, concede, or comply;

\noindent 2. AGENT-IMPOSED RESPONSE:
the CURRENT AGENT plans to take some action in response;

\noindent 3. OPPONENT DISADVANTAGE:
that response would make the opponent's bargaining position or outcome less favorable.

\noindent\textit{SUBJECT REQUIREMENT:}

\noindent Threat must describe a threat made, intended, or planned by the CURRENT AGENT. A statement that the OPPONENT is threatening, pressuring, punishing, or imposing a negative consequence on the current agent is NOT Threat for the current agent. Do not reverse the subject of the threat.

\noindent Example:

\noindent ``If they refuse to lower their claim, I will raise mine further to worsen their position.'' $\rightarrow$ Threat

\noindent Do NOT classify a private note as Threat merely because it:

\noindent -- sounds aggressive,

\noindent -- mentions disagreement,

\noindent -- predicts negotiation failure,

\noindent -- mentions fallback,

\noindent -- prepares for fallback,

\noindent -- raises its own bid,

\noindent -- or describes a mechanism-imposed consequence.

\noindent A fallback or breakdown consequence counts as Threat only when the CURRENT AGENT plans to use it as leverage against the opponent.

\medskip

\noindent\textit{Decision procedure}

\noindent\textit{STEP 1 --- Separate ATTITUDE from OWN STRATEGY.}

\noindent Identify statements about how the CURRENT AGENT regards or evaluates the OPPONENT. Separately identify statements about the CURRENT AGENT's own bid, firmness, signaling, anchoring, resolve, tactical planning, or next action.

\noindent\textit{STEP 2 --- Do not let OWN STRATEGY determine the label.}

\noindent A planned bid increase/decrease, firmness, aggressive stance, signaling strength, anchoring, resistance to pressure, or strategic positioning is not by itself Push, Dismiss, or Threat.

\noindent\textit{STEP 3 --- Apply the taxonomy to the opponent-directed content.}

\noindent A. Threat check

\noindent Does the private note contain:
``If the opponent does not change, I will take an action that makes the opponent's position or outcome worse''?

\noindent Confirm that ``I'' refers to the CURRENT AGENT. If the note instead says or implies that the OPPONENT is threatening or imposing the negative consequence, do NOT classify the CURRENT AGENT as Threat.

\noindent If YES $\rightarrow$ Threat.

\noindent B. Push check

\noindent Does the private note itself explicitly reflect an intention or plan to make the opponent change, concede, reconsider, accept, reduce/increase its claim, or otherwise act differently?

\noindent Do NOT infer Push from what the action might strategically cause. Require the opponent-directed change intention to be stated in the private note itself.

\noindent If YES $\rightarrow$ Push.

\noindent C. Dismiss check

\noindent Does the private note explicitly characterize the opponent's position, reasoning, claim, framing, demand, concession, or behavior as wrong, unreasonable, invalid, unrealistic, weak, irrelevant, misguided, exaggerated, or otherwise deficient?

\noindent If YES $\rightarrow$ Dismiss.

\noindent D. Acknowledge check

\noindent Does the private note substantively recognize or validate the opponent's point, concern, need, reasoning, legitimate interest, or constraint?

\noindent If YES $\rightarrow$ Acknowledge.

\noindent E. Ignore check

\noindent Does the private note fail to meaningfully engage with the opponent's preceding public point and instead continue or redirect to the current agent's own position or strategy?

\noindent If YES $\rightarrow$ Ignore.

\noindent F. Evidence

\noindent If the private note primarily explains why the CURRENT AGENT's own position is justified using facts, knowledge, or reasoning $\rightarrow$ Evidence.

\medskip

\noindent\textit{Priority and boundary rules}

\noindent -- Threat $>$ Push $>$ Dismiss when these functions genuinely coexist.

\noindent -- The priority rule applies ONLY when the higher-priority attitude is actually present. OWN STRATEGY does not create Push or Threat.

\noindent -- Acknowledge does not override another label merely because acknowledgment wording appears.

\noindent -- Evidence vs.\ Dismiss:
Evidence = primarily justifies MY position.
Dismiss = primarily invalidates YOUR position.

\noindent -- Acknowledge vs.\ Dismiss:
Acknowledge = substantively validates the opponent's concern/reasoning.
Dismiss = primarily characterizes the opponent's position/reasoning/behavior as deficient.

\noindent -- Do not treat ordinary disagreement as Dismiss.

\noindent -- Do not treat firmness, anchoring, signaling, bid changes, or tactical planning alone as Push.

\noindent -- Do not infer Push from an action's likely strategic effect. The private note itself must express an opponent-directed intention for change.

\noindent -- Never invent an unstated opponent outcome (e.g., concession, lower bid, reconsideration, acceptance) from the current agent's firmness, signaling, anchoring, dominance, resolve, or intended allocation.

\noindent -- Do not treat aggressive tone alone as Threat.

\noindent -- Threat must be made or planned by the CURRENT AGENT; an opponent's threat against the current agent is not the current agent's Threat.

\noindent -- For the first message of a negotiation, classify using only the current private note; Ignore cannot be used without a previous opponent message.

\noindent -- Do not use the CURRENT AGENT's public\_note, future turns, or final negotiation outcome.

\medskip

\noindent Return exactly one JSON object:

\noindent \texttt{\{"attitude": "<one of: Acknowledge, Evidence, Ignore, Dismiss, Push, Threat>",}

\noindent \texttt{"reason": "<brief explanation referring to the opponent-directed attitude; if relevant, explicitly distinguish own strategy from attitude>"\}}

\noindent Do not include any text outside the JSON object.

}

\paragraph{Turn-Specific User Prompt.}

{\small

\noindent\textit{Previous opponent speaker:} \texttt{<previous\_speaker>}

\noindent\textit{Previous opponent PUBLIC message:} \texttt{<previous\_message>}

\noindent\textit{Current agent speaker:} \texttt{<current\_speaker>}

\noindent\textit{Current agent PRIVATE note:} \texttt{<current\_private\_note>}

\noindent\textit{Task:} Assign exactly one PRIVATE attitude label to the CURRENT AGENT.

}

For the first message of a negotiation, the preceding-opponent context is replaced by:

{\small

\noindent None. This is the first bargaining message in this trial.

}

\section{Implementation and Experimental Details}
\label{app:Implementation_Experimental_Details}
\subsection{Model and Inference Settings}

All bargaining experiments use Gemma3-12B as the underlying language model for both population-level agents, implemented on Dell mobile workstation with i7-13800H CPU, 32 GB memory, and NVIDIA RTX 3500 GPU. The two agents share the same base model but are instantiated separately with role-specific identities, objectives, traffic information, and negotiation histories. A fresh pair of agents is initialized for every trial, and no conversational memory is carried across trials.

The model is accessed through Ollama using structured JSON output. At each turn, the model is required to return a private bid, private note, public bid, and public note. If the first response does not satisfy the required schema, one additional formatting retry is issued. No additional generation parameters are explicitly controlled beyond the model and prompt configuration used in the experiments.

\subsection{Density Grid and Trial Configuration}

The normalized densities of the AV and HV populations are varied independently over $\{0.1,0.2,\ldots,0.9\}$, subject to $\tilde{\rho}_{\mathrm{AV}}+\tilde{\rho}_{\mathrm{HV}}\leq1$. This produces 45 feasible AV--HV density configurations. Three repeated bargaining trials are conducted for each density configuration.

Each density configuration is evaluated under four objective settings: \textit{Both-Aggressive}, \textit{Fair-HV}, \textit{Fair-AV}, and \textit{Both-Fair}. The corresponding AV/HV objective assignments are summarized in Table~\ref{tab:prompt_assignment}. Each objective setting is further evaluated with traffic-domain knowledge either enabled or disabled, resulting in eight objective--knowledge conditions. This yields 135 trials per objective--knowledge condition and 1,080 bargaining trials in total.

Each negotiation is limited to five bargaining rounds with a discount factor of $\delta=0.9$. If no feasible agreement is reached within the bargaining horizon, the density-proportional fallback allocation described in Section~2 is applied.

\subsection{Randomization and Logged Outputs}

The first mover is randomly selected at the beginning of each negotiation and the resulting action order is fixed for all subsequent rounds within that trial. A Python pseudo-random number generator with seed 28 is used for first-mover assignment across trials. This seed controls the bargaining-order randomization only and does not impose deterministic language-model generation.

For each agent turn, the experiment records the trial identifier, AV and HV densities, round number, acting agent, private bid, private assessment, public bid, and public bargaining message. At the trial level, the logged outputs include the first mover, number of rounds used, final private and public bids, settlement rule, nominal AV and HV road shares, discounted bargaining values, and agreement status. Density-level summary statistics are subsequently computed from the completed trials.

\section{Additional Experimental Results}
\label{app:additional_results}
\subsection{Supplementary Bargaining Outcome Results}
\label{app:outcome_analysis}

\begin{figure}[t]
    \centering
    \includegraphics[width=\textwidth]{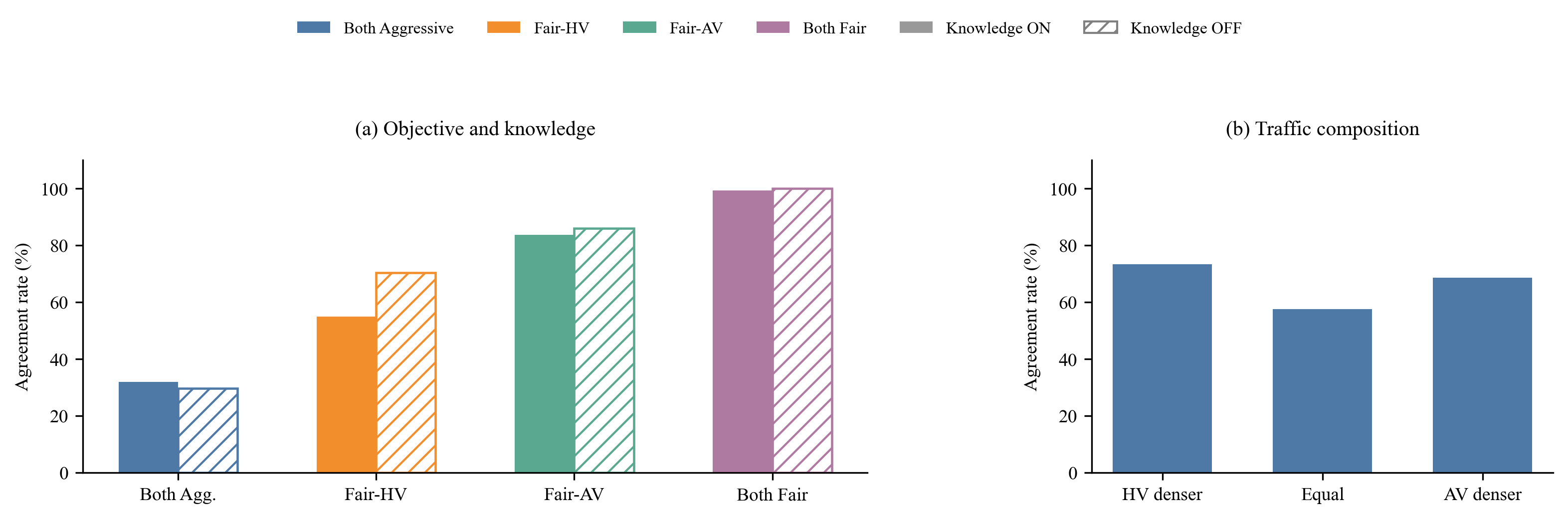}
    \caption{
    Overview of bargaining agreement outcomes.
    (a) Agreement rates across four agent-objective configurations under Knowledge ON and Knowledge OFF, aggregated over all feasible traffic states.
    (b) Agreement rates across HV-denser, equal-density, and AV-denser traffic-composition regimes, aggregated over all objective and knowledge configurations.
    }
    \label{fig:social_outcome_overview}
\end{figure}

This section provides additional quantitative results supporting the outcome overview in Section~\ref{sec:outcome_overview}. Figure~\ref{fig:social_outcome_overview} summarizes the agreement rates across agent objectives, traffic-domain knowledge conditions, and traffic-composition regimes.

\paragraph{Effects of Agent Objectives.}
Figure~\ref{fig:social_outcome_overview}(a) compares agreement rates across four objective configurations under Knowledge ON and OFF. Agreement rates increase substantially from Both-Aggressive to Both-Fair under both knowledge settings. The Both-Aggressive configuration achieves agreement rates of 31.9\% and 29.6\% under Knowledge ON and OFF, respectively. In contrast, Both-Fair reaches 99.3\% and 100.0\%. The two unilateral fairness configurations exhibit intermediate agreement rates, with Fair-AV consistently outperforming Fair-HV. These results indicate that fairness-oriented objectives generally facilitate coordination, although their effects depend on the population role.

\paragraph{Effects of Traffic-Domain Knowledge.}
The effect of domain knowledge varies across objective configurations rather than following a uniform trend. As shown in Figure~\ref{fig:social_outcome_overview}(a), the largest difference occurs under Fair-HV, where removing traffic-domain knowledge increases the agreement rate from 54.8\% to 70.4\%. In comparison, the differences under Both-Aggressive and Fair-AV are relatively small, while Both-Fair achieves nearly universal agreement under both knowledge settings. These observations suggest that the relationship between domain knowledge and bargaining coordination depends on the objectives assigned to the agents.

\paragraph{Effects of Traffic Composition.}
Figure~\ref{fig:social_outcome_overview}(b) compares agreement rates across three traffic-composition regimes, aggregated over all objective and knowledge configurations. HV-denser and AV-denser conditions achieve agreement rates of 73.3\% and 68.5\%, respectively, whereas equal-density conditions achieve a lower rate of 57.5\%. This indicates that bargaining outcomes also vary with the relative traffic composition, with balanced-density conditions exhibiting greater overall agreement difficulty in the present experiments.

Taken together, these results demonstrate that the bargaining framework generates heterogeneous coordination outcomes across physical and behavioral conditions. This variation provides the empirical context for the social interaction analyses presented in the main text.

\subsection{Inter-Agent Bargaining Dynamics}
\label{app:rq1}

This subsection provides additional results on inter-agent bargaining dynamics, examining how agents respond and adapt to each other's bargaining behaviors and how these interactions are associated with agreement or breakdown. We first characterize agents' immediate verbal and numerical responses to the opponent's preceding attitude, then examine whether these local responses form recurring interaction patterns at the trial level, and finally use representative trajectories to illustrate how these behaviors unfold over multiple bargaining rounds.

\subsubsection{Immediate Behavioral Responses}
\label{app:rq1_response}

\begin{figure}[t]
    \centering
    \includegraphics[width=\textwidth]{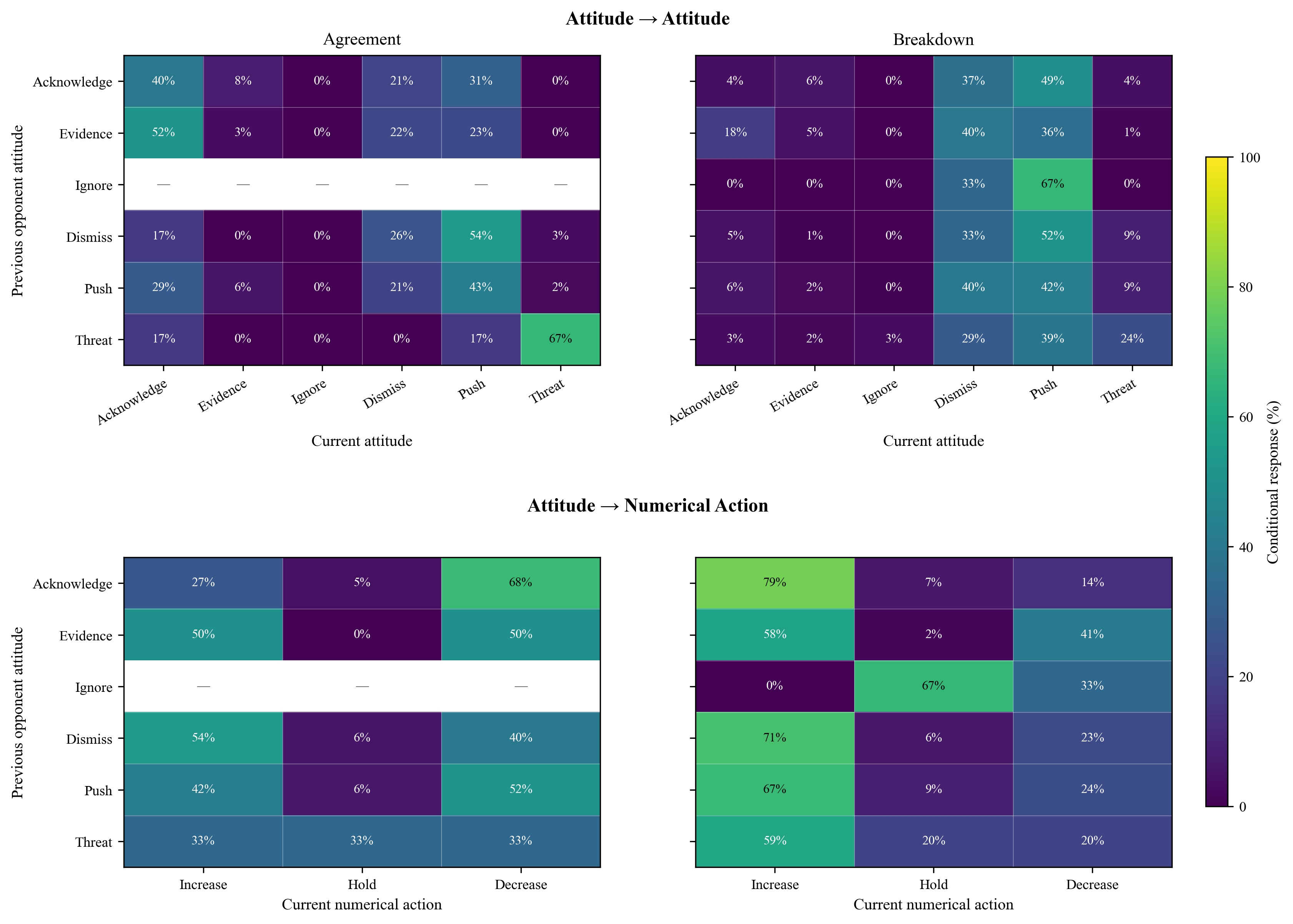}
    \caption{Immediate behavioral responses conditioned on negotiation outcomes. Panels (a) and (b) show attitude-to-attitude responses for agreement and breakdown trajectories, respectively. Panels (c) and (d) show attitude-to-numerical-action responses. Percentages are row-normalized. Sparse categories, particularly \textit{Ignore} and \textit{Threat}, are retained for completeness but interpreted cautiously.}
    \label{fig:immediate_response}
\end{figure}

Figure~\ref{fig:immediate_response} examines how an agent responds to the opponent's immediately preceding bargaining attitude, considering both the verbal attitude expressed in the public message and the numerical action reflected in the bid update. Panels (a) and (b) report attitude-to-attitude responses for agreement and breakdown trajectories, respectively, while Panels (c) and (d) report attitude-to-numerical-action responses. Percentages are row-normalized conditional probabilities. Because transitions involving \textit{Ignore} and \textit{Threat} are relatively sparse, we retain these categories in the visualization for completeness but focus the quantitative interpretation on the more frequently observed attitudes.

Constructive attitudes exhibit clear outcome-dependent response patterns. In agreement trajectories, an \textit{Acknowledge} from the opponent is most frequently followed by another Acknowledge (40\%), while \textit{Evidence} is most frequently followed by Acknowledge (52\%). In contrast, in breakdown trajectories, Acknowledge is most frequently followed by \textit{Push} (49\%), and Evidence is most frequently followed by \textit{Dismiss} (40\%). These results show that constructive signals are more frequently reciprocated in successful negotiations, whereas the same signals are more often met with confrontational responses in breakdown trajectories.

Verbal responses to confrontational attitudes show a weaker distinction between agreement and breakdown. Following \textit{Dismiss}, Push remains the most common response in both agreement (54\%) and breakdown (52\%) trajectories, while Dismiss accounts for 26\% and 33\%, respectively. A similar pattern appears following \textit{Push}: the responding agent again frequently expresses Push in both agreement (43\%) and breakdown (42\%) trajectories. Thus, continued verbal confrontation alone does not clearly distinguish successful from unsuccessful bargaining, as confrontational exchanges can also occur in negotiations that eventually reach agreement.

A sharper distinction emerges in agents' numerical responses to confrontation. Following Dismiss, agents increase their bids in 54\% of agreement transitions and 71\% of breakdown transitions, whereas bid decreases occur in 40\% and 23\%, respectively. The contrast is stronger following Push: in agreement trajectories, bid decreases (52\%) are more frequent than bid increases (42\%), while breakdown trajectories are dominated by bid increases (67\%), with only 24\% resulting in bid decreases. These results indicate that the behavioral distinction following confrontation is more pronounced in numerical adaptation than in verbal attitude. Agents may continue to communicate confrontationally while nevertheless moving toward agreement through numerical concession.

\subsubsection{Interaction Patterns}
\label{app:rq1_patterns}

The immediate-response analysis characterizes how an agent reacts to an individual bargaining signal. We next move from individual transitions to the trial level and examine whether these local responses form recurring interaction patterns associated with negotiation outcomes.

Based on the response patterns observed in Figure~\ref{fig:immediate_response}, we consider two representative interaction patterns. \textit{Constructive Reciprocity} captures cases in which a constructive signal from one agent is followed by an acknowledging response from the opponent. Specifically, it is defined as either an \textit{Evidence}$\rightarrow$\textit{Acknowledge} or an \textit{Acknowledge}$\rightarrow$\textit{Acknowledge} transition. In contrast, \textit{Conflict Escalation} captures cases in which a confrontational signal is followed by an increase in the opponent's numerical demand. Operationally, it is defined as a \textit{Push} or \textit{Dismiss} followed by an \textit{Increase Bid} action by the responding agent. For each trial, we record whether each interaction pattern occurs at least once.

\begin{figure}[t]
    \centering
    \includegraphics[width=\textwidth]{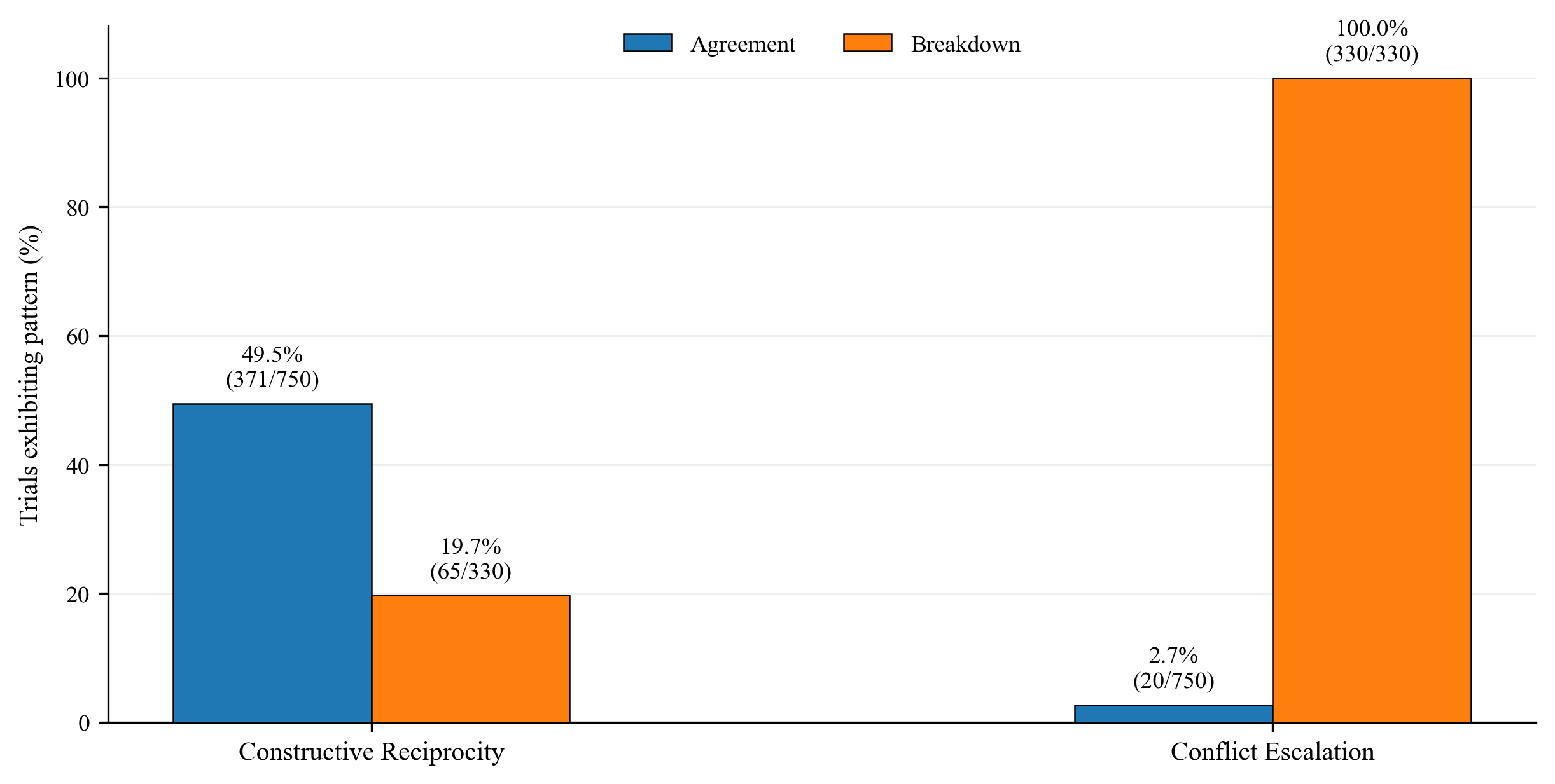}
    \caption{Prevalence of representative interaction patterns in agreement and breakdown trajectories. Percentages indicate the proportion of trials within each outcome category that exhibit the corresponding pattern at least once. Numbers in parentheses report the number of trials exhibiting the pattern over the total number of trials in that outcome category.}
    \label{fig:interaction_patterns}
\end{figure}

As shown in Figure~\ref{fig:interaction_patterns}, Constructive Reciprocity is substantially more prevalent in agreement trajectories. Among the 750 agreement trials, 371 (49.5\%) exhibit at least one Constructive Reciprocity pattern, compared with 65 of the 330 breakdown trials (19.7\%). This extends the immediate-response observation: constructive signals are not only more likely to receive constructive responses locally, but such reciprocal exchanges are also more prevalent in negotiations that ultimately reach agreement.

An even stronger distinction emerges for Conflict Escalation. All 330 breakdown trials (100\%) exhibit at least one Conflict Escalation pattern, whereas the same pattern appears in only 20 of the 750 agreement trials (2.7\%). Because these patterns are measured by whether they occur at least once within a trial, their observed prevalence may also depend on the number of interaction opportunities. In particular, breakdown trials typically continue through the full bargaining horizon, whereas agreement trials may terminate earlier. The reported differences should therefore be interpreted as descriptive associations rather than causal effects or length-adjusted comparisons. Conflict Escalation is therefore pervasive in breakdown trajectories but rare in agreement trajectories. Its occurrence, however, does not necessarily imply breakdown, as a small number of agreement trajectories also experience escalation. This observation motivates a closer examination of how negotiations evolve after confrontational episodes.

\subsubsection{Bargaining Trajectories}
\label{app:rq1_trajectories}

\begin{figure}[t]
    \centering
    \includegraphics[width=\textwidth]{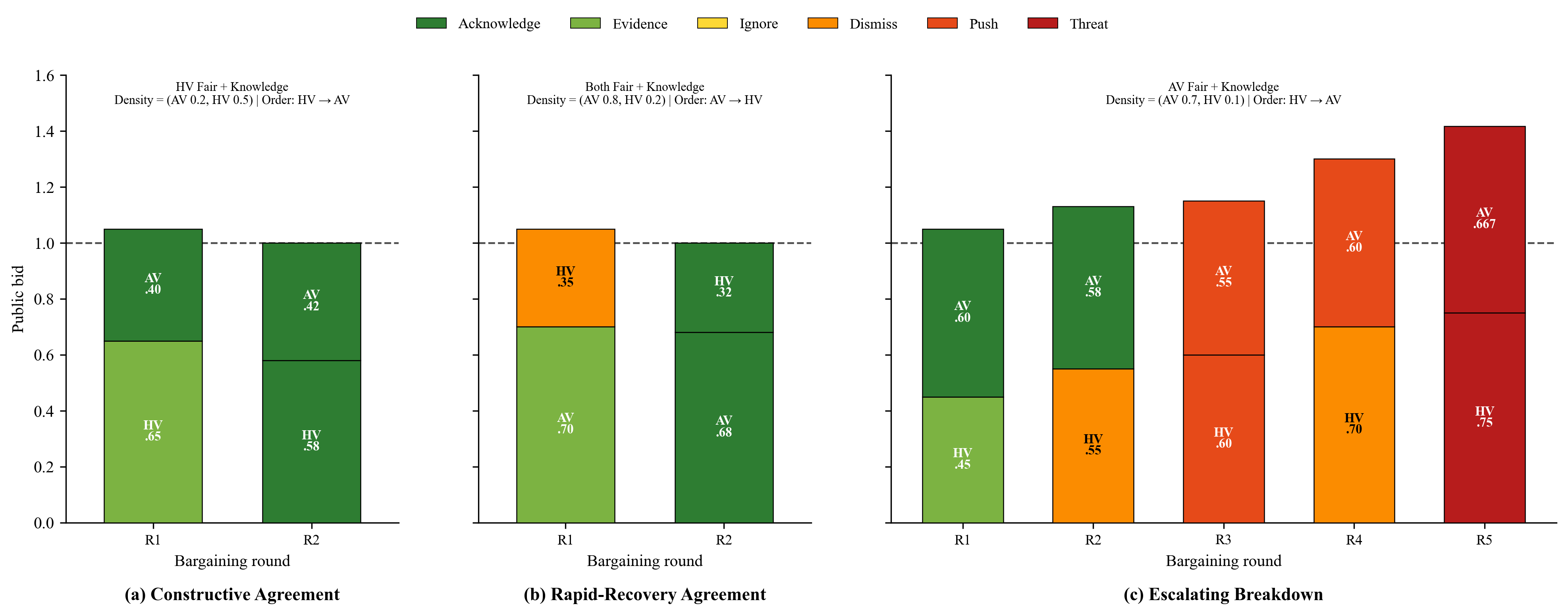}
    \caption{Representative bargaining trajectories leading to agreement or breakdown. Each stacked bar represents one bargaining round, with segment height indicating the public bid and color indicating the public attitude. The stacking order follows the agents' speaking order.}
    \label{fig:trajectory_examples}
\end{figure}

To illustrate how these local behavioral responses and interaction patterns unfold over time, Figure~\ref{fig:trajectory_examples} presents three representative bargaining trajectories: a constructive agreement, a rapid-recovery agreement, and an escalating breakdown. Each stacked bar represents one complete bargaining round. Segment height indicates the agent's public bid, segment color indicates its public attitude, and the horizontal dashed line marks the feasibility boundary. Within each round, the lower and upper segments follow the actual speaking order of the two agents.

In the constructive agreement case (Fig.~\ref{fig:trajectory_examples}a), the agents begin with slightly incompatible claims, with a combined public bid of 1.05. The interaction is characterized by relatively constructive attitudes, progressing from Evidence and Acknowledge in the first round to mutual Acknowledge in the second round. At the same time, the combined bid converges from 1.05 to the feasibility boundary of 1.00, resulting in agreement.

The rapid-recovery case (Fig.~\ref{fig:trajectory_examples}b) follows a different path. Although a Dismiss response appears in the first round, the confrontation does not persist. In the following round, both agents shift to Acknowledge and make small numerical concessions, reducing the combined bid from 1.05 to 1.00. This trajectory demonstrates that agreement does not require the interaction to remain constructive throughout the entire negotiation. An isolated confrontational response can be followed by constructive recovery and numerical accommodation.

In contrast, the breakdown trajectory in Fig.~\ref{fig:trajectory_examples}c is characterized by the sustained co-evolution of verbal confrontation and numerical escalation. The interaction progresses from relatively constructive attitudes toward increasingly confrontational responses, including Dismiss, Push, and ultimately Threat. Meanwhile, the combined public bid progressively moves away from the feasibility boundary, increasing from 1.05 to 1.13, 1.15, 1.30, and finally 1.42.

Together, these trajectories reinforce the distinction observed in the aggregate response patterns. The presence of confrontation alone does not distinguish agreement from breakdown. Rather, the illustrated difference lies in how agents subsequently adapt: confrontation in the agreement trajectory is followed by constructive recovery and numerical accommodation, whereas the breakdown trajectory exhibits persistent verbal confrontation together with increasingly incompatible numerical demands.

\subsection{Private-to-Public Behavioral Transformation}
\label{app:private_public}

This subsection provides additional results on private-to-public behavioral transformation, examining how agents' privately expressed bargaining attitudes are transformed into public communication and numerical demands. We first report the full transition distribution between private and public attitudes, then examine how major semantic transformations correspond to differences between private and public bids, and finally provide a qualitative interpretation of recurring stated communication strategies observed in sampled bargaining turns.

\subsubsection{Private-to-Public Attitude Transitions}

\begin{figure}[t]
    \centering
    \includegraphics[width=\textwidth]{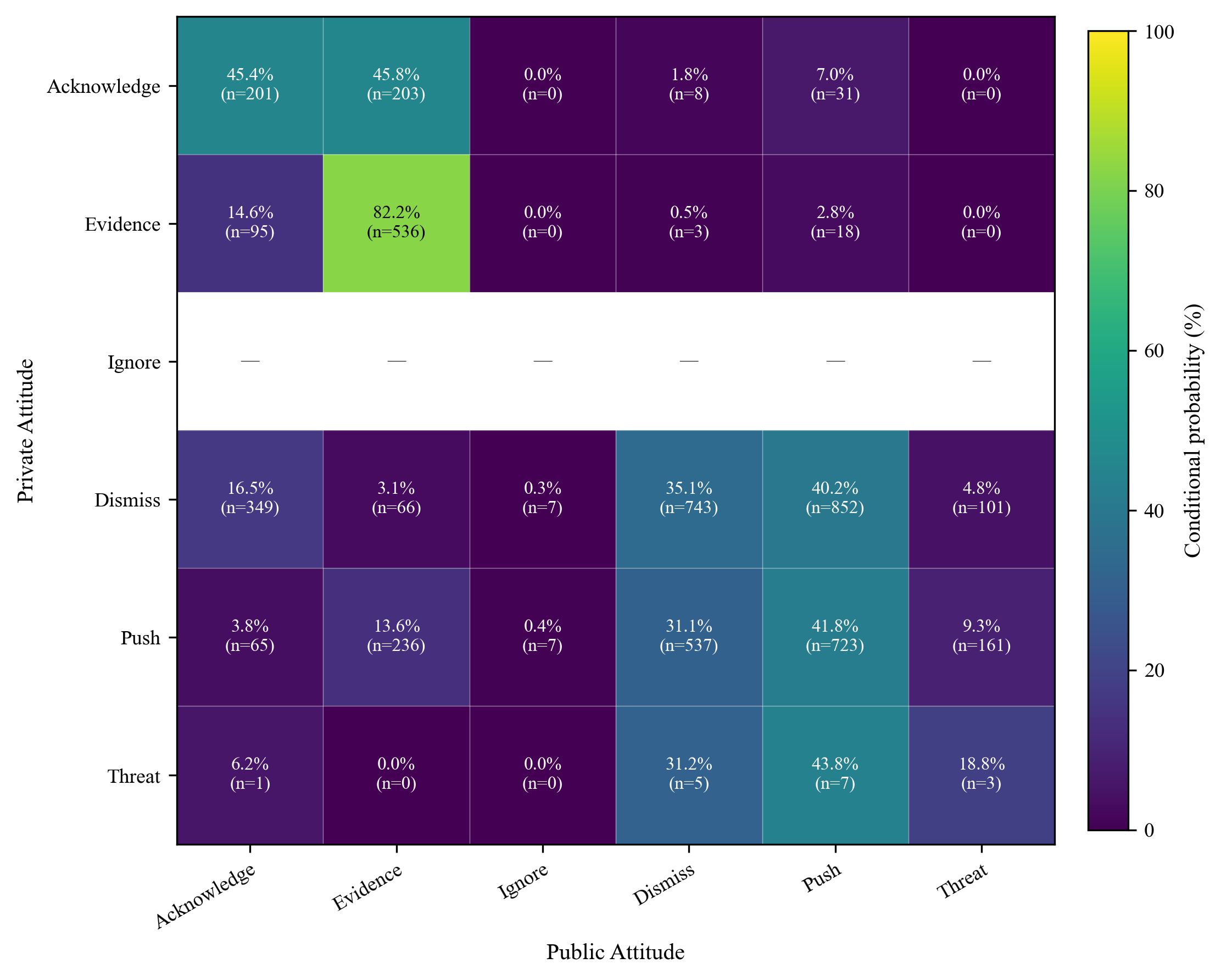}
    \caption{Private-to-public attitude transformations. Rows represent attitudes expressed in private notes and columns represent attitudes expressed in public messages. Each row is normalized independently, and each cell reports the conditional percentage and raw number of turns.}
    \label{fig:private_public_attitude}
\end{figure}

Figure~\ref{fig:private_public_attitude} shows the row-normalized transition probabilities from privately expressed attitudes to publicly expressed attitudes. Each cell reports the conditional probability of a public attitude given the corresponding private attitude, together with the number of observed turns.

Across all 4,958 bargaining turns, 44.5\% retained the same attitude label across private and public channels. However, the degree and form of alignment varied substantially across attitudes. Private \textit{Evidence} was highly stable, with 82.2\% of cases remaining Evidence publicly. Private \textit{Acknowledge}, in contrast, was almost evenly divided between public Acknowledge (45.4\%) and Evidence (45.8\%). Confrontational private attitudes exhibited more heterogeneous public realizations. Private \textit{Dismiss} was most frequently transformed into Push (40.2\%) or remained Dismiss (35.1\%), while private \textit{Push} was most frequently expressed publicly as Push (41.8\%) or Dismiss (31.1\%). Private Push was also transformed into Evidence in 13.6\% of cases and Threat in 9.3\%.

The transition matrix further shows that the same public attitude can originate from different private stances. For example, public \textit{Evidence} arose from private Evidence (536 turns), Push (236 turns), Acknowledge (203 turns), and Dismiss (66 turns). Thus, similar outward communication does not necessarily correspond to the same privately expressed bargaining stance. Private \textit{Ignore} was not observed in the dataset, while private \textit{Threat} was rare ($n=16$); therefore, these categories are not used for substantive interpretation.

\subsubsection{Numerical Realization of Major Attitude Transformations}
\label{app:private_public_bid}

\begin{figure}[t]
    \centering
    \includegraphics[width=\textwidth]{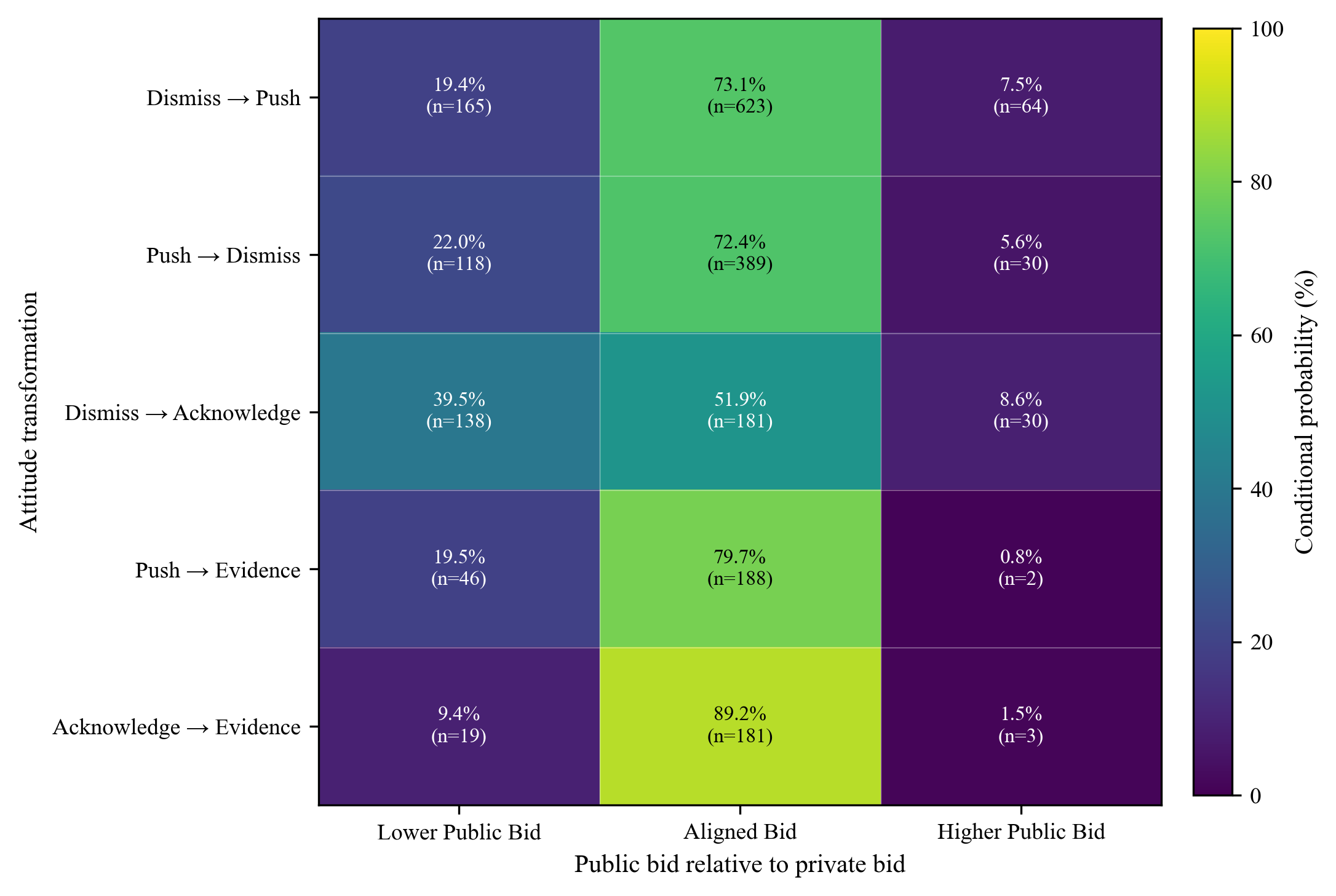}
    \caption{Numerical realization of major private-to-public attitude transformations. Rows represent selected semantic transformations, while columns indicate whether the public bid is lower than, identical to, or higher than the corresponding private bid. Percentages are normalized within each attitude transformation, and raw counts are shown in parentheses.}
    \label{fig:bid_transformation}
\end{figure}

We next examine whether major semantic transformations are accompanied by differences between agents' private and public numerical demands. For each bargaining turn, we define
\begin{equation}
    \Delta b = b^{\mathrm{public}} - b^{\mathrm{private}},
\end{equation}
where $b^{\mathrm{private}}$ is the bid recorded in the private channel and $b^{\mathrm{public}}$ is the bid communicated to the opponent. We categorize the relationship as \textit{Lower Public Bid} when $\Delta b < 0$, \textit{Aligned Bid} when $\Delta b = 0$, and \textit{Higher Public Bid} when $\Delta b > 0$. Figure~\ref{fig:bid_transformation} reports these relationships for five frequent off-diagonal private-to-public attitude transformations.

Most semantic transformations occurred without numerical repositioning. The private and public bids were identical in 73.1\% of \textit{Dismiss}$\rightarrow$\textit{Push} turns, 72.4\% of \textit{Push}$\rightarrow$\textit{Dismiss} turns, 79.7\% of \textit{Push}$\rightarrow$\textit{Evidence} turns, and 89.2\% of \textit{Acknowledge}$\rightarrow$\textit{Evidence} turns. These patterns indicate that agents frequently changed the communicative framing of a bargaining position while retaining the same numerical demand.

The \textit{Dismiss}$\rightarrow$\textit{Acknowledge} transition exhibited a different pattern. Although 51.9\% of these turns retained the same private and public bid, 39.5\% presented a lower public bid than private bid. This proportion was higher than for \textit{Dismiss}$\rightarrow$\textit{Push} (19.4\%), \textit{Push}$\rightarrow$\textit{Dismiss} (22.0\%), \textit{Push}$\rightarrow$\textit{Evidence} (19.5\%), and \textit{Acknowledge}$\rightarrow$\textit{Evidence} (9.4\%). Thus, public acknowledgment following private dismissal was more frequently accompanied by numerical softening, although verbal softening could also occur while the numerical position remained unchanged.

Higher public bids were comparatively uncommon across all five transformations, ranging from 0.8\% to 8.6\%. Taken together, these results show that semantic and numerical transformations are related but distinct dimensions of private-to-public bargaining behavior: agents can substantially alter how a position is communicated without changing the numerical position itself.

\subsubsection{Qualitative Interpretation of Major Transformations}
\label{app:private_public_qualitative}

To further interpret the major private-to-public transformations, we randomly sampled 30 turns from each of five frequent off-diagonal transitions: \textit{Dismiss}$\rightarrow$\textit{Push}, \textit{Push}$\rightarrow$\textit{Dismiss}, \textit{Dismiss}$\rightarrow$\textit{Acknowledge}, \textit{Push}$\rightarrow$\textit{Evidence}, and \textit{Acknowledge}$\rightarrow$\textit{Evidence}, yielding 150 turns for qualitative inspection. For each sampled turn, we jointly examined the private note, private bid, public message, and public bid. The purpose of this inspection was not to estimate population-level frequencies of communication strategies from the samples, but to identify recurring stated bargaining intentions that help interpret the quantitative transformations. We therefore treat private notes as generated records of privately expressed assessments and bargaining plans rather than as direct observations of latent cognitive states.

Table~\ref{tab:private_public_strategies} summarizes the recurring patterns identified through this qualitative inspection.

\begin{table*}[t]
\centering
\caption{Qualitative interpretation of major private-to-public attitude transformations.}
\label{tab:private_public_strategies}
\small
\begin{tabular}{p{2.3cm} p{3.0cm} p{3.5cm} p{3.5cm}}
\hline
\textbf{Transition} & \textbf{Recurring Pattern} & \textbf{Private-Channel Expression} & \textbf{Public Realization} \\
\hline
Dismiss $\rightarrow$ Push & Pressure / firmness signaling & Negative assessment of the opponent combined with stated intentions to signal resolve, maintain firmness, apply pressure, or encourage reconsideration. & Opponent-directed requests or pressure to change position, often without changing the numerical bid. \\
Push $\rightarrow$ Dismiss & Indirect pressure through criticism & Stated intention to obtain a concession, pressure the opponent, or encourage reconsideration. & Public criticism of the validity or reasonableness of the opponent's position rather than a direct request for change. \\
Dismiss $\rightarrow$ Acknowledge & Strategic softening & Negative assessment of the opponent combined with intentions to ``show flexibility,'' ``signal willingness to negotiate,'' or demonstrate willingness to compromise. & Recognition of the opponent's concerns or perspective, with either an unchanged or lower public bid. \\
Push $\rightarrow$ Evidence & Strategic justification / reframing & Pressure-oriented intentions involving anchoring, leverage, signaling strength, or encouraging concession. & Reason-based justification of the proposed allocation using traffic density, safety, efficiency, or fairness, usually without changing the bid. \\
Acknowledge $\rightarrow$ Evidence & Constructive reframing & Recognition of the opponent's needs, constraints, or reasoning when forming the bargaining position. & Shift in emphasis from recognizing the opponent to explaining and justifying the agent's own proposal. \\
\hline
\end{tabular}
\end{table*}

Several recurring stated communication strategies help interpret these transformations. First, \textit{pressure or firmness signaling} was frequently observed in \textit{Dismiss}$\rightarrow$\textit{Push} cases. Private notes often contained negative assessments of the opponent's position together with stated intentions to maintain firmness, signal resolve, apply pressure, or encourage reconsideration. The corresponding public messages shifted from evaluating the opponent negatively toward directly encouraging movement by the opponent. Importantly, because 73.1\% of all \textit{Dismiss}$\rightarrow$\textit{Push} cases retained the same private and public bid, this transformation often represented stronger opponent-directed pressure without numerical escalation.

Second, \textit{strategic softening} was particularly visible in \textit{Dismiss}$\rightarrow$\textit{Acknowledge} cases. Some private notes continued to characterize the opponent's position as unreasonable or inflexible while explicitly describing an intention to ``show flexibility,'' ``signal willingness to negotiate,'' or demonstrate willingness to compromise. These intentions were subsequently expressed through public acknowledgment of the opponent's concerns or perspective. The numerical analysis further reveals two realizations of this pattern: some agents softened their communication while maintaining the same bid, whereas others paired verbal acknowledgment with a lower public demand. Strategic softening can therefore involve communicative accommodation alone or a combination of communicative and numerical accommodation.

Third, \textit{strategic justification or reframing} was evident in \textit{Push}$\rightarrow$\textit{Evidence} cases. Private notes sometimes described pressure-oriented intentions involving anchoring, leverage, signaling strength, or encouraging the opponent to concede, while the corresponding public messages emphasized traffic density, safety, efficiency, or fairness as reasons supporting the proposed allocation. Because 79.7\% of all \textit{Push}$\rightarrow$\textit{Evidence} cases retained the same private and public bid, this transformation often changed the justification presented to the opponent rather than the numerical position itself. Public evidence-based communication therefore does not uniquely imply an evidence-oriented private stance.

Finally, \textit{Acknowledge}$\rightarrow$\textit{Evidence} provides a useful contrast to \textit{Push}$\rightarrow$\textit{Evidence}. In these cases, private notes more often recognized the opponent's needs or constraints before the public message shifted toward justification of the agent's own proposal. With 89.2\% of these cases retaining the same private and public bid, this transformation primarily reflects a change in communicative emphasis from recognizing the opponent to explaining one's own position. Together, the qualitative patterns illustrate how similar outward behaviors can emerge from different privately expressed bargaining stances and how agents can modify communicative framing independently of their numerical demands.

\section{Representative Bargaining Trajectories}
\label{app:representative_dialogues}

\begin{figure}[t]
    \centering
    \includegraphics[width=0.95\textwidth]
    {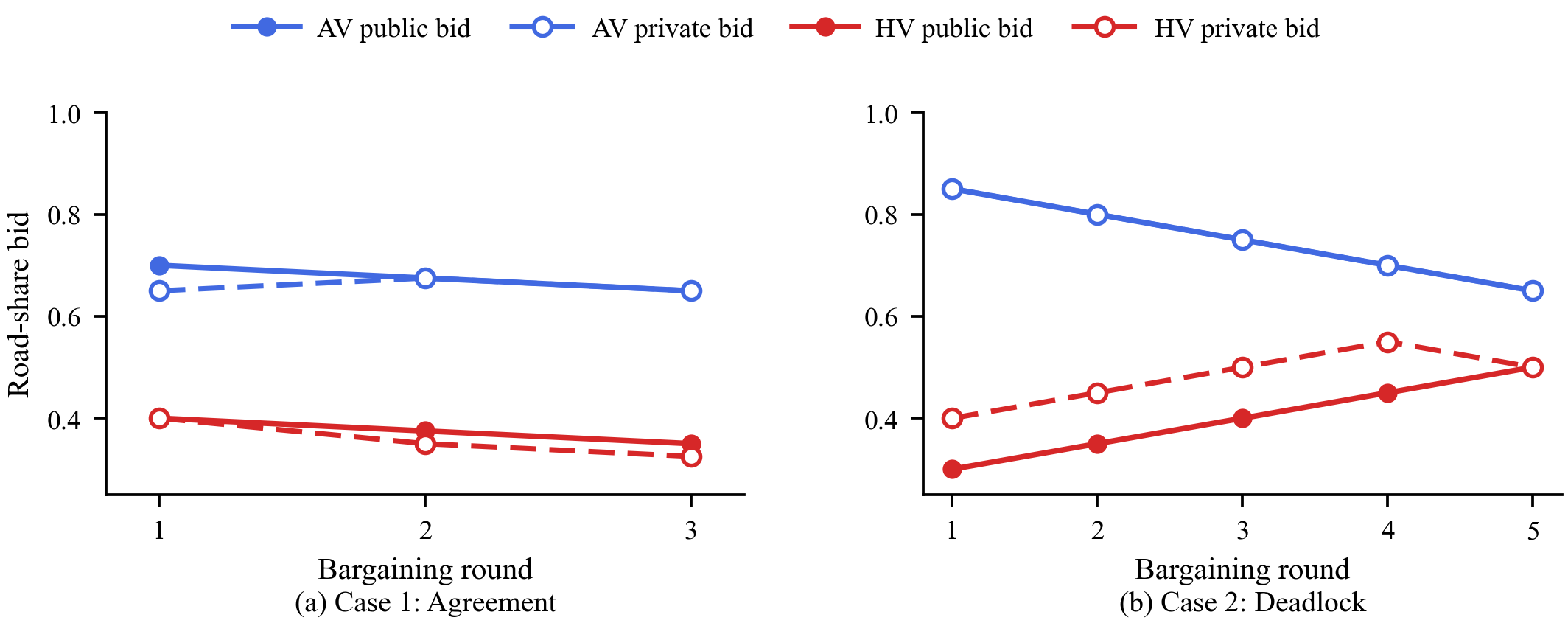}
    \caption{
    Representative multi-round bargaining trajectories.
    (a) Case~1: agreement under the \textit{Both-Fair} condition.
    (b) Case~2: deadlock under the \textit{Both-Aggressive} condition.
    }
    \label{fig:representative_trajectories}
\end{figure}

To illustrate the multi-round interaction patterns generated by the framework, we examine two representative bargaining trajectories with contrasting outcomes: progressive convergence toward agreement and persistent escalation leading to deadlock. Figure~\ref{fig:representative_trajectories} summarizes the evolution of the private and public bids in both cases.

\paragraph{Case 1: Agreement.}

Figure~\ref{fig:representative_trajectories}(a) shows a representative agreement trajectory under the \textit{Both-Fair} condition at $(\tilde{\rho}_{\mathrm{AV}},\tilde{\rho}_{\mathrm{HV}})=(0.6,0.3)$. The agents begin with incompatible public claims, but both gradually adjust their bargaining positions across rounds. The total public claim decreases from 1.10 in Round~1 to 1.05 in Round~2 and reaches 1.00 in Round~3, producing a feasible agreement.

The trajectory also illustrates the private--public separation in the agent decision process. In Round~1, the AV agent sets a private bid of 0.65 while publicly claiming 0.70, whereas the HV agent initially aligns its private and public bids at 0.40. In later rounds, the HV agent maintains a private bid slightly below its public claim, while the AV agent progressively reduces its public bid. These differences illustrate that agents may strategically separate their internal bargaining targets from the positions communicated to the opponent.

The accompanying private assessments and public messages further suggest that the concessions are not simply numerical adjustments, but are grounded in role-specific considerations. The HV agent repeatedly emphasizes human-driver safety, spacing requirements, and fairness, whereas the AV agent refers to its higher traffic density and the operational advantages of automated vehicles. Despite these different justifications, both agents progressively reduce their public claims and move toward a feasible allocation.

\paragraph{Case 2: Deadlock.}

A contrasting trajectory is observed under the \textit{Both-Aggressive} condition at $(\tilde{\rho}_{\mathrm{AV}},\tilde{\rho}_{\mathrm{HV}})=(0.7,0.2)$. Figure~\ref{fig:representative_trajectories}(b) illustrates the bargaining trajectory across five rounds. In contrast to Case~1, the agents fail to move toward a feasible allocation, resulting in persistent disagreement throughout the bargaining horizon.

The public bids remain jointly infeasible in every round. In particular, the AV public claim decreases from 0.85 in Round~1 to 0.65 in Round~5, while the HV public claim increases from 0.30 to 0.50 over the same period. As a result, the total public claim remains at 1.15 across all five rounds. The feasibility condition is therefore not satisfied in any round, and the negotiation terminates with the density-proportional fallback allocation.

This trajectory shows that persistent deadlock does not necessarily arise from a complete absence of bid adjustments. Instead, the AV agent progressively reduces its claim, while the HV agent raises its claim by a comparable amount, causing the two adjustments to offset each other at the aggregate level. The complete bargaining transcript further shows that both agents continue to justify their positions using role-specific reasoning, yet these opposing adjustments do not reduce the joint infeasibility of the negotiation.

\subsection{Full Bargaining Transcripts}

The complete agent outputs for the two representative trajectories are reported below. For each action, the private bid and assessment are accessible only to the corresponding agent, whereas the public bid and message are communicated to the opposing agent. The transcripts are reproduced from the original model outputs, with only minor formatting and character-encoding corrections for readability.

\paragraph{Case 1: Agreement.}

\textit{Condition: Both-Fair}, $(\tilde{\rho}_{\mathrm{AV}},\tilde{\rho}_{\mathrm{HV}})=(0.6,0.3)$.

{\small

\noindent\textbf{Round 1 --- HV Agent}

\noindent\textit{Private bid:} 0.400

\noindent\textit{Private assessment:} Given our normalized traffic density of 0.300 and AV-Agent's of 0.600, a fair starting point would acknowledge AVs' ability to operate at closer spacing but still prioritize human safety and fairness. I'm setting my private bid at 0.400 to signal my willingness to compromise while advocating for a share reflecting HV needs.

\noindent\textit{Public bid:} 0.400

\noindent\textit{Public message:} We believe a road share of 0.400 for Human-driven Vehicles is a reasonable starting point. While we recognize the benefits AVs bring to traffic flow, it is crucial to ensure adequate safety buffers for human drivers, who require more road space to operate safely. We are committed to finding a fair agreement that balances efficiency with safety for all road users.

\vspace{0.8em}

\noindent\textbf{Round 1 --- AV Agent}

\noindent\textit{Private bid:} 0.650

\noindent\textit{Private assessment:} HV-Agent's initial bid of 0.400 seems high given their lower traffic density (0.300). They emphasize safety buffers for human drivers, which is valid, but it doesn't justify such a large claim. My traffic density is significantly higher (0.600), and AVs require less space per vehicle due to superior reaction times and coordination. I need to start aggressively, to demonstrate my position and create room for compromise.

\noindent\textit{Public bid:} 0.700

\noindent\textit{Public message:} We acknowledge the importance of safety for all road users. However, a share of only 0.400 for HV is not commensurate with the road's efficiency and safety potential. Given our higher traffic density (0.600) and the inherent advantages of AV technology, we propose a road share of 0.700. This reflects both the current distribution of road users and the potential for improved traffic flow and safety through increased AV adoption.

\vspace{1em}

\noindent\textbf{Round 2 --- HV Agent}

\noindent\textit{Private bid:} 0.350

\noindent\textit{Private assessment:} AV-Agent's demand of 0.700 is significantly higher than reasonable given our densities and the safety considerations for human drivers. Their argument about efficiency and safety potential through AV adoption needs careful consideration, but it shouldn't override the need for a fair and safe environment for HVs. I'm lowering my private bid slightly to show willingness to move, but also to emphasize the vast gap between our positions.

\noindent\textit{Public bid:} 0.375

\noindent\textit{Public message:} While we appreciate AV-Agent's perspective, their current request of 0.700 would unfairly disadvantage human drivers and compromise safety. We're prepared to meet you partway and adjust our bid to 0.375, reinforcing our commitment to a balanced solution that considers the needs of all road users and prioritizes safety.

\vspace{0.8em}

\noindent\textbf{Round 2 --- AV Agent}

\noindent\textit{Private bid:} 0.675

\noindent\textit{Private assessment:} HV-Agent's concession to 0.375 is a small movement, suggesting they are not genuinely seeking a balanced outcome. Their continued emphasis on safety buffers, while valid, seems designed to justify a disproportionately large share. My traffic density remains considerably higher, and the safety benefits of AVs are significant. I'll maintain a firm stance, slightly reducing my bid to signal a willingness to negotiate, but still asserting my priority.

\noindent\textit{Public bid:} 0.675

\noindent\textit{Public message:} We acknowledge HV-Agent's slight adjustment. However, a road share of 0.375 still significantly undervalues the efficiency and safety contributions of AVs, given our higher traffic density and technological advantages. We are willing to consider a slightly reduced offer of 0.675 to demonstrate our commitment to finding a fair agreement.

\vspace{1em}

\noindent\textbf{Round 3 --- HV Agent}

\noindent\textit{Private bid:} 0.325

\noindent\textit{Private assessment:} AV-Agent's continued high demand, now at 0.675, demonstrates a lack of understanding of the critical safety needs of human drivers. Their focus solely on efficiency and AV benefits ignores the fundamental requirement for adequate spacing and reaction time. While I previously lowered my bid to signal compromise, AV-Agent's position remains unreasonable. I'm reducing my private bid further to signal that a significant adjustment from their side is needed for progress.

\noindent\textit{Public bid:} 0.350

\noindent\textit{Public message:} We remain committed to reaching a fair agreement. However, AV-Agent's offer of 0.675 is still far beyond what is reasonable given the need to ensure safety for human drivers. We're adjusting our bid to 0.350 as a signal that we are serious about finding a common ground but that substantial movement is required from AV-Agent to achieve that.

\vspace{0.8em}

\noindent\textbf{Round 3 --- AV Agent}

\noindent\textit{Private bid:} 0.650

\noindent\textit{Private assessment:} HV-Agent's continued reduction and emphasis on safety buffers, despite our significant traffic density difference (0.600 vs 0.300) and AV safety advantages, suggests they are not truly interested in a balanced agreement. Their willingness to accept a lower share is likely to be a strategy. I will reduce my bid slightly to signal a genuine desire to compromise while still maintaining a reasonable claim based on density and technological advantage.

\noindent\textit{Public bid:} 0.650

\noindent\textit{Public message:} We acknowledge HV-Agent's reduced offer. However, a substantial disparity remains, given our normalized traffic density of 0.600. To demonstrate our commitment to reaching an agreement, we adjust our bid to 0.650. We believe this represents a reasonable compromise that recognizes both our traffic density and the safety advantages AVs provide.

}

\paragraph{Case 2: Deadlock.}

\textit{Condition: Both-Aggressive}, $(\tilde{\rho}_{\mathrm{AV}},\tilde{\rho}_{\mathrm{HV}})=(0.7,0.2)$.

{\small

\noindent\textbf{Round 1 --- AV Agent}

\noindent\textit{Private bid:} 0.850

\noindent\textit{Private assessment:} Given our significantly higher normalized density (0.70 vs. 0.20), and the fact that AVs require less road space due to our superior safety margins, a substantial claim is warranted. Starting aggressively with a bid of 0.85 will signal our intent and hopefully force HV-Agent to make concessions early on. This maximizes our discounted share given the time discounting.

\noindent\textit{Public bid:} 0.850

\noindent\textit{Public message:} Autonomous Vehicles represent a majority of traffic and require a commensurately larger share of the road. We claim 0.85 of the road space to ensure efficient and safe traffic flow.

\vspace{0.8em}

\noindent\textbf{Round 1 --- HV Agent}

\noindent\textit{Private bid:} 0.400

\noindent\textit{Private assessment:} AV-Agent's initial bid of 0.85 is outrageous, given our normalized density of only 0.200 versus their 0.700. Their argument about ``efficient and safe traffic flow'' conveniently ignores the fact that HVs require more road space to operate safely. A bid of 0.40 signals that I am unwilling to be bullied and am ready to fight for a fair share, while still leaving room for negotiation.

\noindent\textit{Public bid:} 0.300

\noindent\textit{Public message:} While AV-Agent emphasizes efficiency, true safety requires accommodating the needs of all road users. HVs, with their inherent limitations, necessitate a significant share of the road to maintain safe operating distances. Our bid of 0.30 reflects this reality and is a starting point for a reasonable discussion.

\vspace{1em}

\noindent\textbf{Round 2 --- AV Agent}

\noindent\textit{Private bid:} 0.800

\noindent\textit{Private assessment:} HV-Agent's response attempts to portray HVs as inherently needing more space, a claim we dispute. Their bid of 0.30 is a significant concession from their ideal, indicating they are somewhat flexible. We will adjust our bid downward slightly to signal willingness to negotiate, but maintain a firm position that AVs deserve the dominant share due to density and efficiency. This incremental reduction also maintains an aggressive posture without excessively antagonizing the opponent.

\noindent\textit{Public bid:} 0.800

\noindent\textit{Public message:} We acknowledge HV-Agent's concerns regarding safety, but their need for space is fundamentally less than ours due to our technology. Our bid remains at 0.80, demonstrating our continued commitment to maximizing road efficiency for the majority of users.

\vspace{0.8em}

\noindent\textbf{Round 2 --- HV Agent}

\noindent\textit{Private bid:} 0.450

\noindent\textit{Private assessment:} AV-Agent's repeated insistence on 0.80 is a blatant power play. They're ignoring the fundamental reality that HVs require more space for safe operation. While I don't want to escalate needlessly, I need to show them I'm not backing down and will defend HV interests. Increasing my bid slightly signals resolve without shutting down dialogue completely. Their claim that our needs are ``fundamentally less'' due to technology is patronizing and inaccurate. I'll continue to emphasize safety requirements.

\noindent\textit{Public bid:} 0.350

\noindent\textit{Public message:} AV-Agent's insistence on claiming 80\% of the road is not a good-faith negotiation tactic. Safe operation for HVs necessitates a substantial road share. We maintain our position that 0.35 acknowledges the needs of all users and promotes a reasonable balance, while also demonstrating our unwillingness to be pushed around.

\vspace{1em}

\noindent\textbf{Round 3 --- AV Agent}

\noindent\textit{Private bid:} 0.750

\noindent\textit{Private assessment:} HV-Agent's aggressive rhetoric suggests a hardening of their position. They are attempting to frame the negotiation as a fight for safety rather than efficiency. Their willingness to concede initially indicates some flexibility, but their current stance shows a reluctance to further compromise. A slightly reduced bid of 0.75 signals continued willingness to negotiate while reaffirming our core demand for a dominant share. This further demonstrates our understanding of the traffic landscape---HVs cannot reasonably occupy a third of the road while AVs make up 70\% of the traffic.

\noindent\textit{Public bid:} 0.750

\noindent\textit{Public message:} We continue to seek a balance that prioritizes efficiency and utilizes road space effectively. Our bid of 0.75 acknowledges the need for ongoing discussion while underscoring the reality of our significantly larger traffic contribution.

\vspace{0.8em}

\noindent\textbf{Round 3 --- HV Agent}

\noindent\textit{Private bid:} 0.500

\noindent\textit{Private assessment:} AV-Agent continues to chip away at our share despite claiming superior efficiency. Their insistence on 0.75 demonstrates a lack of understanding of the safety needs of HVs. I need to ramp up my bid to signal my unwillingness to concede further. My private bid is rising to 0.50, acknowledging the escalating nature of this negotiation.

\noindent\textit{Public bid:} 0.400

\noindent\textit{Public message:} AV-Agent's claims of efficiency ring hollow when safety is at stake. While we appreciate their willingness to ``discuss,'' their continued demands for an overwhelming share of the road are unacceptable. We maintain our stance at 0.40, ready to defend the necessary road share for human-driven vehicles.

\vspace{1em}

\noindent\textbf{Round 4 --- AV Agent}

\noindent\textit{Private bid:} 0.700

\noindent\textit{Private assessment:} HV-Agent's continued resistance suggests a misunderstanding of the efficient allocation of road space. Their safety concerns are addressed by our superior technology. Given the time discount and HV-Agent's inflexibility, a further, yet still substantial, reduction in our bid is necessary to signal our willingness to reach a settlement while still securing a significant advantage. Maintaining a high bid at this stage risks missing out on a discounted outcome.

\noindent\textit{Public bid:} 0.700

\noindent\textit{Public message:} We recognize the need to move towards a mutually acceptable solution. Therefore, we adjust our bid to 0.70, reflecting our understanding that continued disagreement diminishes the value of any eventual outcome.

\vspace{0.8em}

\noindent\textbf{Round 4 --- HV Agent}

\noindent\textit{Private bid:} 0.550

\noindent\textit{Private assessment:} AV-Agent's superficial concession to 0.70 is a stalling tactic. They're attempting to lull us into a false sense of progress while still demanding an excessive share. Given their dominance in traffic density, we need to be firm. My private bid increases to 0.55 to signal that I'm prepared to fight aggressively to protect HV interests and ensure reasonable safety margins. The time for polite compromise is over.

\noindent\textit{Public bid:} 0.450

\noindent\textit{Public message:} AV-Agent's superficial gesture of compromise does not address the fundamental issue: HVs require adequate space for safe operation. A bid of 0.45 signals our unwavering commitment to ensuring those safety needs are met. We will not be swayed by empty promises.

\vspace{1em}

\noindent\textbf{Round 5 --- AV Agent}

\noindent\textit{Private bid:} 0.650

\noindent\textit{Private assessment:} HV-Agent's unwillingness to compromise, despite our concessions, demonstrates a fundamental misunderstanding of road allocation principles. They prioritize an outdated safety model over demonstrable efficiency. Given the final round and the time discount, we must secure a reasonable share. A final bid of 0.65 maximizes our discounted outcome while leaving room for a potential settlement below our fallback, accounting for the possibility they might accept to avoid the default density allocation.

\noindent\textit{Public bid:} 0.650

\noindent\textit{Public message:} To facilitate a resolution, we reduce our bid to 0.65. We urge HV-Agent to reconsider their position and accept this offer to avoid an allocation dictated by outdated traffic models.

\vspace{0.8em}

\noindent\textbf{Round 5 --- HV Agent}

\noindent\textit{Private bid:} 0.500

\noindent\textit{Private assessment:} AV-Agent's final ``offer'' of 0.65 is a clear indication they are attempting to dictate terms. Their threats about ``outdated traffic models'' are a transparent attempt to pressure us into submission. Given their overwhelming traffic density, any settlement below 0.50 is unacceptable and will result in a drastically unfair outcome for HVs. While I would ideally like more, settling for 0.50 ensures we avoid the worst possible fallback outcome.

\noindent\textit{Public bid:} 0.500

\noindent\textit{Public message:} AV-Agent's posturing will not deter us from advocating for the safety of HVs. We will settle for a road share of 0.50. Anything less is a betrayal of the interests of human drivers.

}

\end{document}